\documentclass[reprint, bibnotes, amsmath,amssymb, aps, prx, floatfix, showkeys,superscriptaddress]{revtex4-2}

\usepackage{graphicx}
\usepackage{dcolumn}
\usepackage{bm}
\usepackage{braket}
\usepackage{xcolor}

\usepackage[text={7in,10in},centering]{geometry}

\begin{document}

\preprint{APS/123-QED}

\title{Multistability and Noise-Induced Transitions \\in Dispersively-Coupled Nonlinear Nanomechanical Modes}

\newcommand{\BU}{ Department of Mechanical Engineering, Division of Materials Science and Engineering, and the Photonics Center, Boston University, Boston, Massachusetts 02215, United States}

\newcommand{\VT}{Department of Mechanical Engineering, Virginia Tech, Blacksburg, Virginia 24061, United States}

\newcommand{\GC}{Department of Physics, Gordon College, Wenham, Massachusetts 01984, United States}

\newcommand{\SUNUM}{SUNUM, Nanotechnology Research and Application Center, Sabanci University, Istanbul, 34956, Turkey}

\newcommand{\Sabanci}{Faculty of Engineering and Natural Sciences, Sabanci University, Istanbul, 34956, Turkey}

\newcommand{\Bilkent}{Department of Mechanical Engineering, Bilkent University, Ankara, 06800, Turkey}

\newcommand{\UNAM}{National Nanotechnology Research Center (UNAM), Bilkent University, Ankara, 06800, Turkey}

\author{David Allemeier}
\affiliation{\BU}
 \email{davidallemeier@gmail.com}
 
 \author{I. I. Kaya}
\affiliation{\SUNUM}
\affiliation{\Sabanci}

\author{M. S. Hanay}
\affiliation{\Bilkent}
\affiliation{\UNAM}

\author{K. L. Ekinci}
 \email{ekinci@bu.edu}
\affiliation{\BU}

\date{\today}

\begin{abstract}

We study the noisy dynamics of two coupled bistable modes of a nanomechanical beam. When de-coupled, each driven mode obeys the Duffing equation of motion, with a well-defined bistable region in the frequency domain. When both modes are driven, intermodal dispersive coupling emerges due to the amplitude dependence of the modal frequencies and leads to coupled states of the two modes. We map out the dynamics of the system by sweeping the drive frequencies of both modes in the presence of added noise. The system then samples all accessible states at each combination of frequencies, with the probability of each stable state being proportional to its occupancy time at steady state. In the frequency domain, the system exhibits four stable regions---one for each coupled state---which are separated by five curves. These curves are reminiscent of coexistence curves in an equilibrium phase diagram: each curve is defined by robust inter-state transitions, with equal probabilities of finding the system in the two contiguous states. Remarkably, the  curves intersect in two triple points, where the system now transitions between three distinct contiguous states. A physical analogy can be made between this nonequilibrium system and a  multi-phase thermodynamic system, with possible applications in computing, precision sensing, and signal processing.
 
\end{abstract}

\maketitle

\section{Introduction}
\begin{figure*}
    \includegraphics[]{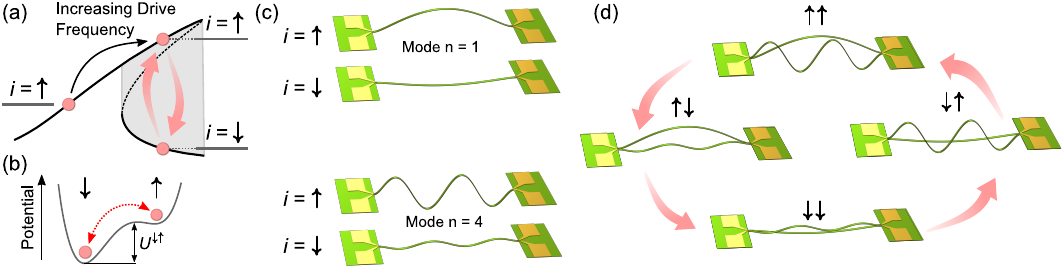}
    \caption{\label{fig:intro} (a) Bistability in a Duffing resonator.  Within a certain frequency range of the resonator response, there are two stable states, labeled $\uparrow$ and $\downarrow$. (b) Intuitive representation of the $\uparrow$ and $\downarrow$ states of the Duffing resonator as minima of the system potential. Random fluctuations can drive the system over the barrier and induce a state change. (c) The two stable states of modes 1 and 4 when driven in the nonlinear regime. The high amplitude upper states are labeled $\uparrow$ and the low amplitude states are labeled $\downarrow$. The states are located on the upper and lower branches of the Duffing response in (a). (d) Four possible coupled states of modes 1 and 4, $\uparrow\uparrow$, $\uparrow\downarrow$, $\downarrow\uparrow$, and $\downarrow\downarrow$, which are defined by the nonlinear states occupied by the two modes. Transitions can potentially occur between each of the four coupled states.}
\end{figure*}

One of the most intriguing phenomena exhibited by a  periodically forced nonlinear system is \textit{multistability}. A multistable nonlinear system can reside in one of several stable oscillation states under a given set of conditions. A classic and well-studied example of multistability occurs in a driven Duffing resonator, which has both a quartic and a harmonic term in its potential. The nonlinearity in this system is characterized by a Duffing coefficient $\alpha$ that quantifies the strength and direction (stiffening or softening) of the nonlinearity \cite{strogatz,pikovsky_synch, nayfeh_nonlinear, lc_nems, roukes_nems}. Above a critical forcing amplitude and frequency, as illustrated in Fig. \ref{fig:intro}a, the response of a driven Duffing resonator bifurcates such that the system can oscillate in one of two distinct stable states---one at high amplitude and in-phase with the drive (denoted $\uparrow$), and the other at low amplitude and out-of-phase with the drive (denoted $\downarrow$) \cite{strogatz, pikovsky_synch, landau, nayfeh_nonlinear}. Under these conditions, the Duffing resonator is \textit{bistable} and exhibits hysteresis, settling in either the $\uparrow$ or $\downarrow$ state according to the initial state of the system.

Intuitively,  the two stable states of a Duffing resonator can be thought of as the minima of a double-well potential (Fig. \ref{fig:intro}b); more rigorously, the states are the extrema of the effective potential of the system in the frame rotating at the drive frequency \cite{dyk_og, dyk_squeezing, dyk_scaling, dyk_quantumstats}. Just like a particle trapped in a double-well potential or coexisting phases in a thermodynamic system, random fluctuations can drive the system over the activation barrier and trigger a transition between the two states \cite{dyk_og, dyk_3d, scheefer_cts, ma_cts, shi_cts, chan_poisson, badzey_SR, venstra_cantiswitching}. The rate of transitions $W^{ij}$ from state $i$ to $j$, which are the $i,j=$ $\uparrow$ or $\downarrow$ states of a bistable Duffing resonator, is governed by Kramers' rule and is typically expressed in terms of the energy $D^i$ of the fluctuations, a characteristic activation barrier $U^{ij}$, and a maximum transition rate $R^{ij}$  \cite{dyk_hfsr, dyk_3d, udani_detEsc}:
\begin{equation}
\label{eq:trans_rate}
    W^{ij} = R^{ij}{\exp}\left(-\frac{U^{ij}}{D^{i}}\right).
\end{equation}
Over  time scales significantly longer than $1/W^{ij}$ and $1/W^{ji}$, the bistable system undergoes many transitions and reaches a non-equilibrium steady-state characterized by the respective occupancy probabilities of the $\uparrow$ and $\downarrow$ states, $P^\uparrow$ and $P^\downarrow$. These probabilities are fully determined for a particular drive amplitude and vary monotonically as a function of drive frequency. For a system with a stiffening nonlinearity ($\alpha>0$), such as the one we study here, at low frequency $P^\uparrow \gg P^\downarrow$ and at high frequency $P^\downarrow \gg P^\uparrow$\cite{dyk_3d,defoort_scaling, chan_barrier, cleland_switching}; in systems with a softening nonlinearity ($\alpha <0$), the opposite is true. In either case, at a particular drive frequency the probability of the two states becomes equal. In analogy to equilibrium thermodynamic phase transitions, this point has been referred to as the \textit{kinetic phase transition} (KPT) point \cite{dyk_hfsr, chan_supernarrow, dyk_quantumstats, roy_optphase}.

The Duffing equation of motion describes the behavior of a wide range of physical systems that are driven nonlinearly, including mechanical resonators \cite{lc_nems,nayfeh_nonlinear}, Josephson junctions \cite{vijay_bifamp}, superconducting resonators \cite{beaulieu_kerrSuper}, Kerr optical cavities \cite{li_opticalSwitching}, and even chemical reactions \cite{epstein_nonlinChem}. In a nanomechanical beam, such as the one shown in Fig. \ref{fig:intro}, each eigenmode, when driven nonlinearly, can be described as a Duffing resonator and exhibit bistability \cite{yuksel, dario_stabilization, chan_switching, cleland_switching, dyk_meso, koz_basins, roukes_nems}. The nonlinearity in this system emerges from the tension induced during high amplitude oscillations that results in an effective stiffening of the beam, making $\alpha > 0$. Since tension is a global parameter shared between all modes, increasing the amplitude of any mode  causes an up-shift in the eigenfrequencies of all the modes of the structure. The resulting interdependence of the mode amplitudes and  eigenfrequencies is known as \textit{dispersive (or reactive) coupling} \cite{lc_nems, matheny_modes, sader_intermodal, westra_modal}.

So far, certain aspects of dispersive interactions between multiple modes in a nanomechanical system have been explored in fundamental studies \cite{matheny_modes, karabalin_chaos, vinante_mixing, westra_modal} and exploited in applications  \cite{sader_intermodal, aravind_coupledLogic, cross_pulling, atakan_coupling, chan_coupled, cross_qnd, dario_stabilization, matheny_synch, roukes_bif, shevyrin_tuning}. Several studies have focused on characterizing the dispersive interactions of two modes within an elastic structure in terms of the coupling constants \cite{sader_intermodal, westra_modal, matheny_modes, shevyrin_tuning}. Intermodal dispersive coupling has also been used to detect changes in modal parameters \cite{atakan_coupling, cross_qnd}, and  stabilize \cite{dario_stabilization} and tune \cite{shevyrin_tuning, lu_dispersion, zhou_modulation} modal frequencies  of a given structure. In general, dispersive coupling can arise between the modes of two distinct, sometimes nominally identical, structures \cite{chan_ising, mahboob_ising}, mediated by elastic or electromagnetic fields. These effects can lead to the exchange of signals and noise between multiple modes  \cite{matheny_synch, cross_pulling, chan_coupled, vinante_mixing, karabalin_chaos, mahboob_ising}.  Furthermore, the physics of these coupled systems  have been mapped onto a reconfigurable asymmetric Ising spin system \cite{chan_ising}, where the dynamical nonlinear states of each oscillator structure become analogous to the spin states of a quantum system.

Despite the above-described body of work, the dynamics of two dispersively-coupled, bistable nonlinear modes  has yet to be fully been elucidated. Fig. \ref{fig:intro}c  shows the complexity of the problem. When both modes $n=1$ and $4$ are driven in their bistable regimes, for instance, the system can be in one of four possible states: $\uparrow\uparrow$, $\uparrow\downarrow$, $\downarrow\uparrow$, and $\downarrow\downarrow$. The large change in amplitude when one of the modes transitions to a different state can dramatically affect the behavior of the other mode. In analogy with the above-mentioned Ising model, this coupling may be positive, where the symmetric $\uparrow\uparrow$ and $\downarrow\downarrow$ states are preferred, or negative, where the asymmetric states are favorable. Furthermore, any change in either the force amplitude or force frequency may change the amplitudes of both modes due to the intermodal coupling, possibly triggering unexpected jumps into different states. These effects may modulate the effective potential of each mode and lead to the breakdown of the  single-mode description in Eq. \ref{eq:trans_rate}, perhaps requiring a formulation based on two interacting particles in a double-well potential. Finally, the one-dimensional KPT picture of a single bistable mode may need to be extended appropriately.

Here, we study this problem of two coupled, bistable nonlinear modes by employing the $n=1$ and $4$ modes of a nanomechanical beam resonator. In particular, we rely on fluctuation-enabled transitions to map the stability of the states of the coupled system in the frequency (detuning) domain.  We identify four distinct stable states of the system; in between these stable regions, we observe  transitions between up to three different states. These transitions allow us to construct two-dimensional KPT curves in the frequency domain that generalize the KPT description of a single bistable mode with two states. The end result is a detailed understanding of the dynamics and inter-state transitions in a system made up of interacting nonlinear subsystems. The paper is organized as follows. In \S\ref{section:experiment}, we describe our experiment. In \S\ref{sec:single-mode}, we show that each mode under study obeys the Duffing equation individually. In \S\ref{sec:coupled-modes}, we turn to the dispersive interactions between two modes. We measure the amplitude maps for the two coupled modes and identify the stable states using noise-enabled transitions; we then provide the KPT curves and discuss state transitions. The final section is reserved for our conclusions. 

\section{Experimental Setup}
\label{section:experiment}
\begin{figure*}
     \includegraphics[]{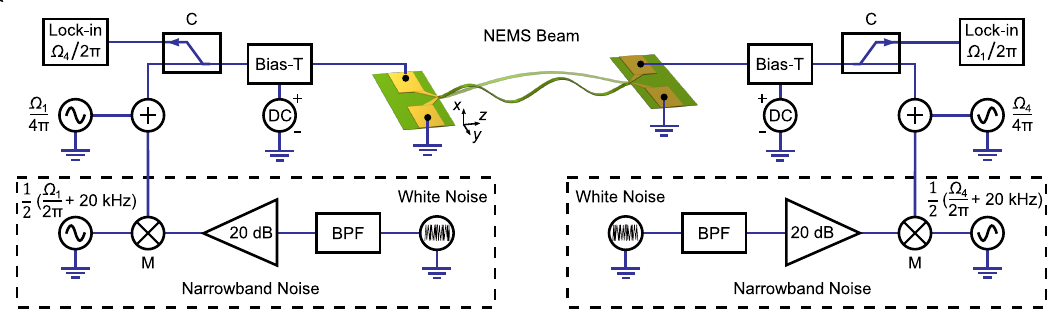}
     \caption{\label{fig:circuit} Circuit diagram and connections to the silicon nitride beam used in the experiments performed on modes 1 and 4 of the beam. The beam has gold electrodes fabricated on both anchors for actuation and detection of its motion. A lock-in amplifier was used to detect the motion of each mode. Acronyms  are defined as follows: NEMS: nanoelectromechanical systems; BPF: band-pass filter (10-50 kHz); M: mixer; C: directional coupler. } 
 \end{figure*}
 
To conduct this experiment, we used a doubly-clamped silicon-nitride (SiN) nanomechanical beam of length $L = 50 ~\rm \mu m$ and cross-section $S = 900 \times 100$ nm$^2$. We studied  the $n=1$ and $4$ out-of-plane eigenmodes, shown in Fig. 2. The beam was kept in a high vacuum of $2 \times 10^{-5}$ Torr to reduce viscous damping. Actuation and detection of the beam motion was achieved through approximately $50 ~\Omega$ gold (Au) nanoresistors deposited at each end of the beam \cite{chaoyang}.  Using the beam dimensions and  SiN material properties of $E \approx 250$ GPa and $\rho = 3000$ kg/m$^3$, we estimated the intrinsic tensile stress to be 10 MPa \cite{monan_scaling, atakan_brownian}. This gave a non-dimensional tension parameter of $U\approx400$. Since $U \gg 1$, the beam dynamics were nearly string-like \cite{monan_scaling}.

We actuated the beam electrothermally using Joule heating at the nanoresistors, which produced a force through the thermal expansion mismatch between the Au resistor and the SiN beam \cite{roukes_nems,monan_scaling}. The transduction of the drive signal from the electrical to the mechanical domains exhibited a quadratic nonlinearity due to the square dependence of the temperature on the applied voltage $V$. This doubled the  frequency of the force relative to the frequency of the sinusoidal voltage and resulted in mixing of the drive voltage components \cite{monan_jap}. To manage these effects, we separated the two drives onto opposite sides of the beam. We then selected the first and fourth modes for study to avoid both the 1:3 internal resonance and the interference between the drive and detection signals due to the quadratic nonlinearity in transduction.

Figure \ref{fig:circuit} shows a block diagram of our actuation and detection circuit. Our measurements were performed using a lock-in amplifier operated in $2\omega$ mode.  The actuation was based on the above-mentioned electrothermal effect, while the detection exploited the piezoresistivity of the Au nanoresistor \cite{chaoyang}. To implement the two mode actuation-detection scheme,  we applied three electrical signal components to each side of the beam: a sinusoidal drive at nearly half the eigenfrequency of mode $n$, which we denote $\Omega_n/2\pi$, where $n=1, 4$; a narrowband noise signal; and a small DC offset. We generated the narrowband noise by bandpass filtering  broadband white noise  in the range of 10-50 kHz; we then up-converted the baseband noise such that one of the mixed noise bands overlapped with the mode eigenfrequency. The small DC offset provided a bias for piezoresistive  detection. We detected each mode on the  side opposite to its drive to avoid any crosstalk.

In separate sets of measurements, we calibrated the  input (electrothermal) and output (piezoresistive) transducers on the beam. The displacement amplitude of the beam was calibrated using a Michelson interferometer in the linear regime \cite{bello_interferometer}. The piezoresistance signal was estimated to stay extremely linear, with an expected error of less than 5\% at the maximum displacement of $250$ nm in the first mode. The force from the electrothermal transducer was calibrated using the bifurcation points of the two modes. The onset of bistability for modes $n=1$ and 4, respectively, occurred at critical amplitudes of $A_{1c}\approx 6.1$ nm and $A_{4c}\approx 2.2$ nm at the critical force values  of $F_{1c} \approx 0.5$ pN and $F_{1c} \approx 6.3$ pN \cite{monan_scaling}. In the rest of the paper, we report the peak amplitudes of (coherent) oscillatory  physical quantities, such as the preceding force and amplitude values, whereas noise will typically be in rms units. All the relevant experimental details are included in the Supplemental Material \cite{myreference}. We summarize the measured and calculated parameters of the two modes in Table \ref{tab:beam_params}.

\begin{table*}
    \begin{tabular}{c|c|c|c|c|c|c|c|c|c}
        Mode $n$ & ${\Omega_n \over 2\pi}$ [MHz] & $k_n$ [N/m] & $M_n$ [pg] & $Q_n$ & $\Gamma_n\over 2\pi$ [Hz] & $A_{nc}$ [nm] & $F_{nc}$ [pN] & $\alpha_{nn} \over 4\pi^2$ [MHz$^2$/nm$^2$] & $\alpha_{nm} \over 4\pi^2$ [MHz$^2$ /nm$^2$]  \\ [0.5ex] \hline \hline 1 & 2.292 & 1.4 & 6.8 & 20,100 & 57 & 6.1 & 0.5 & $1.08 \times 10^{-5}$ & $17.7 \times 10^{-5}$ \\ 
        4 & 9.995 & 26.6 & 6.8 & 11,000 & 454 & 2.2 & 6.3 & $277.1  \times 10^{-5}$ & $17.8 \times 10^{-5}$
    \end{tabular}

    \caption{Summary of measured and calculated beam parameters for modes $n=1$ and $n=4$. The eigenfrequencies $\Omega_n\over 2\pi$ and Duffing coefficients $\alpha_{nn}$ and $\alpha_{nm}$ are directly available from experiments. The masses $M_n$ and spring constants $k_n$ are approximated using $M_n=\rho LS/2$ and $k_n=M_n{\Omega_n}^2$, respectively. The quality factors $Q_n$, critical amplitudes $A_{nc}$, and critical forcings $F_{nc}$ are obtained by fitting to the theoretical bifurcation relations, as discussed in the Supplemental Material \cite{myreference}. }
    \label{tab:beam_params}
\end{table*}
 
\section{Nonlinear Dynamics of Single Nanomechanical Modes}
\label{sec:single-mode}
\subsection{Single-mode Nonlinear Response}

We begin our analysis with the driven nonlinear response of a single mode of the beam. To this end, we write the Euler-Bernoulli beam equation under tension \cite{ginsberg_vibrations,lc_nems},
\begin{equation}
    \label{eq:euler}
    \mu\ddot{u} = -EIu'''' + Tu'',
\end{equation}
where $u = u(z,t)$ is the displacement of the beam at position $z$ and time $t$, $\mu = \rho S$ is the linear mass density, $E$ is the elastic modulus, $I$ is the area moment of inertia, and $T$ is the tension in the beam. Prime and dot respectively denote  derivatives with respect to $z$ and $t$. The tension contains the intrinsic tension  $T_0$  as well as a dynamic term that arises from the elongation of the beam during vibration \cite{lc_nems},
\begin{equation}
    \label{eq:tension}
    T(t) = T_0 + \frac{ES}{2L}\int_0^L \left[u'(z,t)\right]^2 dz.
\end{equation}
The dynamic tension gives rise to the nonlinearity in the system and couples the motion between modes through its dependence on all modal displacements.

\begin{figure*}
    \includegraphics[]{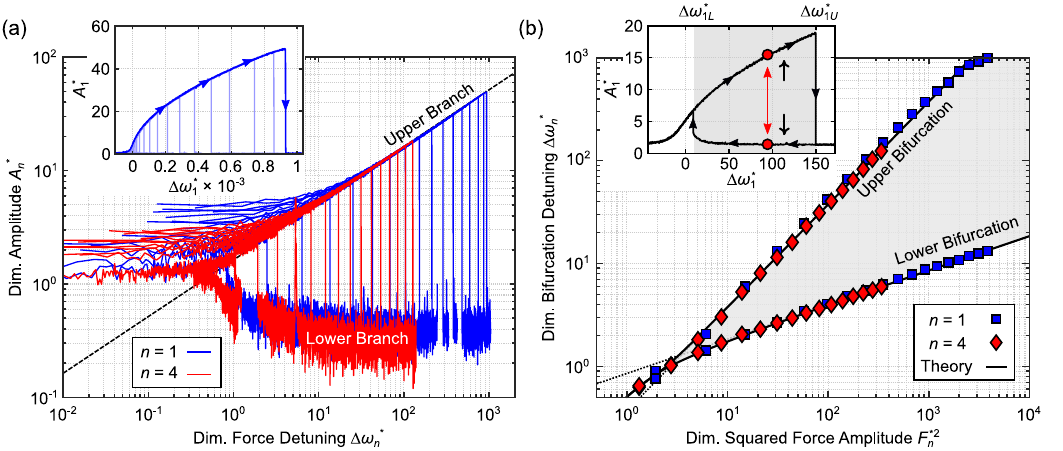}
    \caption{\label{fig:duffing}(a)  Double-logarithmic plot of the forward-swept frequency response of modes 1 and 4 in dimensionless variables. The modes were driven independently at drive powers varying from -23.5 to -4.8 dBm in steps of $\sim1$ dB, resulting in approximate amplitudes of 7 nm to 250 nm for mode $n=1$, and 2 nm to 35 nm for mode $n=4$. These values approximately correspond to $1.5<F_1^*<62$, $1<F_4^*<18$, $2<A_1^*<50$, and $3<A_4^*<19$, respectively. The black dashed line shows the theoretical Duffing backbone, which is the asymptotic approximation in Eq. \ref{eq:amp_approx}. Inset shows the same data for mode 1 in a linear plot. (b) Dimensionless upper and lower bifurcation detunings of modes 1 and 4 as a function of the dimensionless forcing amplitude squared, ${F^*}^2$. The solid lines show the exact theoretical predictions given by Eq. \ref{eq:app_discrim}; the dashed lines show the asymptotic approximations in Eqs. \ref{eq:bif_approx} and 12. The inset is a bi-directional frequency sweep of the $n=1$ mode response and shows the positions of the dimensionless lower and upper bifurcation frequencies, $\Delta \omega_{1L}^*$ and $\Delta \omega_{1U}^*$, respectively, for a fixed drive force of $F_1^* = 30.3$. The states $\uparrow$ and $\downarrow$ are indicated by the dots at an example  drive detuning. Further details can be found in the Supplemental Material \cite{myreference}.  }
\end{figure*}

Following standard steps, we express the beam displacement in mode $n$ as $u(z,t)=x_n(t) \Phi_n(z)$, where $\Phi_n(z)$ is the mode shape of mode $n$ normalized at the largest antinode and $x_n(t)$ is the time-dependent mode amplitude at this position, with $n = 1,2,3\dots$. Each mode is characterized by an eigenfrequency $\Omega_n \over 2\pi$,  a spring constant $k_n$, and a phenomenological damping rate $\Gamma_n$, all directly available from experiments. We add a sinusoidal force term $F_n\cos{\left[\left(\Omega_n + \Delta \omega_{n}\right)t\right]}$ expressed in terms of the detuning $\Delta \omega_n$ of the force from $\Omega_n$. Finally, we reach the Duffing equation of motion for  mode $n$ \cite{lc_nems, ochs_thesis},
\begin{eqnarray}
        \label{eq:eom_final}
\ddot{x}_n + 2\Gamma_n\dot{x}_n + \left({\Omega_{n}}^2 + \alpha_{nn}{x_n}^2 \right)x_n =\\ \nonumber  \frac{F_n}{M_n}\cos{\left[\left(\Omega_n + \Delta \omega_{n}\right)t\right]},
\end{eqnarray}
where $\alpha_{nn}$ is the intramodal Duffing coefficient that connects the mode amplitude to the mode frequency. The solution for the mode amplitude $A_n$ at the forcing frequency can be found as \cite{ochs_thesis,dyk_meso,lc_nems},
\begin{equation}
    \label{eq:amp_resp}
    {A_n}^2 = \frac{{F_n}^2}{{4{{{M_n}^2}{\Omega_n}^2}}\left[\left(\Delta \omega_n - {3 \over 8} {\alpha_{nn} \over \Omega_n}{A_n}^2\right)^2 + {\Gamma_n}^2\right]},
\end{equation}
which is a cubic function of ${A_n}^2$ and captures the Duffing response observed in the inset of Fig. \ref{fig:duffing}a. A corresponding expression can be found for the phase $\phi_n$ \cite{lc_nems}. A derivation of Eq. \ref{eq:amp_resp} consistent with this work may be found in the Appendix.

For a linear resonator, i.e., at small amplitude,   Equation \ref{eq:amp_resp} converges to a  Lorentzian centered at $\Omega_n$, indicating that resonance occurs around the eigenfrequency. In a Duffing resonator, however, the amplitude alters the frequency of the resonance peak. At large amplitude, it is convenient to invert the Duffing amplitude response function  to isolate $\Delta \omega_n$ in Eq. \ref{eq:amp_resp} in terms of the mode amplitude $A_n$ and force $F_n$ as \cite{nayfeh_perturb,nayfeh_nonlinear,ochs_thesis}
\begin{equation}
    \label{eq:amp_inverted}
    \Delta\omega_n = \frac{3}{8}\frac{\alpha_{nn}{A_n}^2}{\Omega_{n}} \pm \left({\frac{{F_n}^2}{4{M_n}^2{A_n}^2} - {\Gamma_n}^2}\right)^{1 \over 2},
\end{equation}
although  the force detuning $\Delta \omega_n$ is typically the independent variable in experiments. At the limit of large amplitude and small dissipation,  Eq. \ref{eq:amp_resp} or \ref{eq:amp_inverted} provides an asymptotic relation for $A_n$,
\begin{equation}
    \label{eq:amp_approx}
    {{A_n}^2 = {8 \over 3}{\Omega_n \over \alpha_{nn}}\Delta \omega_n},
\end{equation}
which indicates that the steady-state displacement becomes independent of the force amplitude and is  solely determined by  $\Delta \omega_n$. This results in the convergence of frequency sweeps conducted at different drive powers at large $\Delta \omega_n$. The approximate relation in Eq. \ref{eq:amp_approx} is used extensively in our analyses below. 

We next define dimensionless variables based on the Duffing response. In Appendix \ref{app:nd}, we nondimensionalize Eq. \ref{eq:eom_final} following Lifshitz and Cross \cite{lc_nems}.  Based on this nondimensionalization, our experimental results are presented in terms of 
\begin{eqnarray}
    \label{eq:nondim}
   \Delta \omega^*_n &=& \frac{\Delta \omega_n}{2\Gamma_n}, \\
   A^*_n &=& A_n{\frac{\sqrt{\alpha_{nn} Q_n}}{\Omega_n}} ,\\ 
    F^*_n &=& F_n {\sqrt{\alpha_{nn}{Q_n}^3}\over {M_n\Omega_n}^3}.
\end{eqnarray}
The variables $\Delta \omega^*_n$, $A^*_n$, and $F^*_n$ represent the dimensionless force detuning, oscillation amplitude, and force amplitude, respectively, with the following physical meanings. The force detuning $\Delta \omega_n$ is expressed in units of the mode linewidth $2\Gamma_n$, such that $\Delta\omega_n^* = {Q_n \Delta\omega_n\over \Omega_n}$. The oscillation amplitude is first nondimensionalized by the factor $\sqrt{\alpha_{nn}}/\Omega_n$, where an amplitude of ${\cal O}(1)$ would result in roughly a doubling of the mode frequency; the  amplitude is further scaled by a factor of $\sqrt{Q_n}$, which corresponds to the limit of weak nonlinearity \cite{lc_nems, nayfeh_perturb}. Similarly, the force is   nondimensionalized by $\sqrt{\alpha_{nn}}/M_n{\Omega_n}^3$ and  ${Q_n}^{3/2}$. This leads to the force, dissipation, and cubic amplitude terms appearing in the same order in the small parameter $\epsilon = {Q_n}^{-1}$ in the dimensionless version of Eq. \ref{eq:eom_final}. Physically, representing the variables in this way ensures that the onset of nonlinearity occurs at the same values of $A^*_n$, $F^*_n$, and $\Delta\omega^*_n$ for multiple modes and allows for direct comparison of the oscillation amplitudes, as we demonstrate in Fig. \ref{fig:duffing}. Further details are provided in the Appendix. 


The experimental Duffing amplitude response of the two modes of our beam   is shown in Fig. \ref{fig:duffing}a. The response was measured by driving each mode separately, and the data are presented in a double-logarithmic plot in dimensionless units, with the corresponding  data for mode 1 shown in the linear plot in the inset.   The frequency response for each mode exhibits two distinct branches, labeled upper branch and lower branch in Fig. \ref{fig:duffing}a. On the upper branch, each mode  approaches the asymptotic amplitude-frequency relation in Eq. \ref{eq:amp_approx}, which is independent of the force amplitude. This asymptote  is sometimes referred to as the Duffing backbone \cite{landau, lc_nems, nayfeh_nonlinear}. We fit the maximum amplitude and detuning at each drive amplitude to Eq. \ref{eq:amp_approx} to calculate the values of the intramodal Duffing coefficients as $\alpha_{11}/4\pi^2 = 1.08\times10^{-5} ~\rm MHz^2/nm^2$ and $\alpha_{44}/4\pi^2 = 277.1\times10^{-5} ~\rm MHz^2/nm^2$ \cite{monan_scaling, myreference}.  Throughout the paper, we will follow the conventions used in this figure and present the mode $n=1$ and $4$ data in blue and red, respectively. 

\subsection{Single-mode Bistability}
\label{sec:bistability}

We return to the full amplitude response in Eq. \ref{eq:amp_resp} to expand our analysis to the lower branch shown in Figure \ref{fig:duffing}a. Eq. \ref{eq:amp_resp} is a cubic function of ${A_n}^2$ and has three complex roots. At low amplitude, the effect of nonlinearity is negligible and only one of the roots is real. The single real solution traces a  Lorentzian, with a slight loss of symmetry as the drive intensity increases. Above a well-defined critical force and drive frequency that define the onset of bistability, the cubic equation jumps from one to three real roots, referred to as a \textit{bifurcation} \cite{strogatz, pikovsky_synch, nayfeh_nonlinear}. Two of these roots represent stable oscillation states while the third is an unstable saddle-node \cite{dyk_og}. The critical values for the force $F_{nc}$ and amplitude $A_{nc}$ for each mode are listed in Table I.

The two stable states can be observed in a bi-directional frequency sweep, such as the one shown in the inset of Fig. \ref{fig:duffing}b. For a fixed drive frequency, we label these states $\uparrow$ and $\downarrow$, indicated with red dots. The $\uparrow$ state lies on the upper branch of the Duffing response and is accessed by sweeping the drive frequency upward from below resonance. The $\uparrow$  state is high amplitude and nearly in-phase with the drive. The $\downarrow$ state is located on the lower branch of the Duffing response and is accessed by sweeping the drive frequency down from above resonance. This state is low amplitude and is out-of-phase with the drive by roughly $\pi$ radians. The upper and lower branches become unstable at well-defined frequencies referred to as the upper and lower bifurcation detunings, $\Delta \omega_U$ and $\Delta \omega_L$, respectively. The amplitude response at these points is discontinuous, and the mode spontaneously jumps to the other branch when swept through these points. The system thus exhibits hysteresis and traces either the upper or lower branch of the Duffing response depending on the initial drive conditions.

Fig. \ref{fig:duffing}b shows the dimensionless   lower and upper bifurcation detunings, $\Delta \omega^*_{nL}$ and $\Delta \omega^*_{nU}$, for both modes  as a function of the dimensionless squared force amplitude ${F^*_n}^2$.  We show in Appendix \ref{sec:app-bif} that, far from the onset of bistability, the bifurcations occur at dimensionless detunings given approximately in terms of the dimensionless force  \cite{defoort_scaling, nayfeh_nonlinear} as
\begin{eqnarray}
        \label{eq:bif_approx}
        \Delta \omega^*_L &=& \sqrt[3]{\frac{81}{128}}{F^*}^{2/3}, \\ \Delta \omega^*_U &=& \frac{3}{8}{F^*}^2.
\end{eqnarray}
Eqs. 11 and 12 are shown by the two dotted lines in Fig. \ref{fig:duffing}b, which coincide with the  continuous curves for large ${F^*_n}^2$. The continuous curves are the exact analytical solutions that can be found in Appendix \ref{sec:app-bif}.  The two bifurcation curves deviate from the approximations in Eq. \ref{eq:bif_approx} and 12 (dotted lines) as the mode approaches the onset of bistability \cite{monan_scaling}, at which point the two curves merge into one. Excellent agreement is observed between the measured bifurcation points and the theory. The frequency range between the two bifurcations for a fixed $F^*$ defines a region where both branches are stable, i.e., the \textit{bistable region}. When the drive force and frequency lie outside this region, the system will be monostable in either the $\uparrow$ or $\downarrow$ state. 

\subsection{State Transitions in a Single Bistable Mode}
\label{sec:trans}
\begin{figure*}
    \includegraphics[]{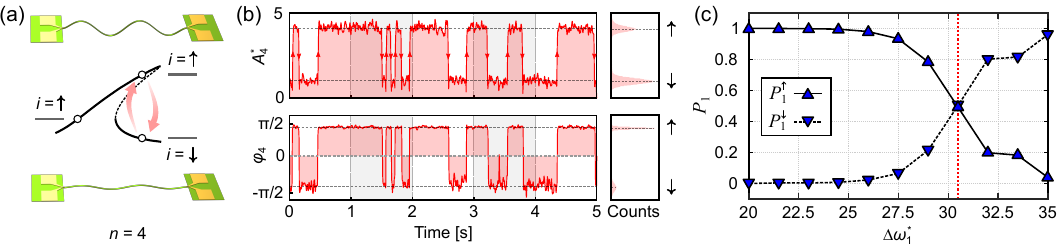}
    \caption{\label{fig:single-switching} (a) The upper and lower states of mode 4. The two states represent the two  solutions on the Duffing response. (b) Transitions between the $\uparrow$ and $\downarrow$ states  of mode 4  in time domain, with the  amplitude and phase of the oscillations switching simultaneously.  The data was taken at a drive power of -6.7 dBm ($F_4^* = 11.7$) and a detuning of $\Delta \omega^*_4=6.5$, corresponding to a steady-state amplitude of $A_4 \approx 8$ nm in the upper state. The applied noise power density was -74.3 dBm/Hz, corresponding to an rms noise amplitude of approximately 0.6 nm. We applied a phase shift of $\approx \pi/2$ to make the phase threshold zero. The histograms show that the fluctuations are Gaussian in both amplitude and phase. The fluctuations in phase are significantly larger in the lower amplitude state due to the fact that phase noise increases with decreasing oscillation amplitude \cite{cleland_noise}. (c) Measurement of the state probabilities for mode 1 at a drive power of -10 dBm ($F_1^* = 30.3$) and noise power density of -67.9 dBm/Hz, corresponding to coherent and noise amplitudes of $A_1 \approx 30$ nm and 4 nm, respectively. At this drive power, the dimensionless lower and upper bifurcation frequencies are $\Delta \omega_{1L}^* = 8.3$ and $\Delta \omega_{1U}^* = 344$ respectively. The location of the KPT is indicated with a dashed red line.  Each data point was calculated from a 120 s time-domain trace. 
    }
\end{figure*}

In the bistable regime of each mode, thermal and applied noise cause fluctuations about the stable points that occasionally exceed the potential barrier and trigger a  transition from one state to the other (Fig. 4a) \cite{dyk_hfsr, venstra_cantiswitching, chan_poisson, koz_basins, chan_switching, buks_amp, steeneken_graphene, venstra_cantiswitching, scheefer_cts,ma_cts,shi_cts}. In Fig. \ref{fig:single-switching}b, we show the state transitions in mode 4 when driven at constant amplitude and frequency in the presence of artificial Gaussian noise. The two stable states, $\uparrow$ and $\downarrow$, appear as flat regions that are separated in amplitude and phase, where we have shifted the phase by $\pi/2$ so that $\phi^\uparrow \approx \pi/2$ and $\phi^\downarrow \approx -\pi/2$. The artificial noise gives rise to two Gaussian profiles in the histograms centered on each stable point as the system undergoes rapid transitions from one plateau to the other.

The rate of transitions $W^{ij}_n$ between the two states of a given mode $n$ is governed by Eq. \ref{eq:trans_rate} and depends on the activation barriers for each transition. For relatively low noise levels, the activation barriers are solely functions of the detuning of the force from the bifurcation points presented in Fig. \ref{fig:duffing}b \cite{dyk_hfsr, defoort_scaling, chan_barrier, chan_switching}. Over long time scales $\Delta t\gg 1/W^{ij}_n$, the system transitions back and forth between the two states and reaches a non-equilibrium steady-state condition quantified by the state probabilities $P_n^\uparrow$ and $P_n^\downarrow$. These probabilities follow standard relations for a two-level non-equilibrium system, i.e., $P^\uparrow_n + P^\downarrow_n = 1$, and $P_n^i = {\tau_n^i / (\tau_n^i + \tau_n^j)}$,  where $\tau^i_n$ is the average lifetime of  state $i$. The state probabilities and lifetimes obey the master equation $ P_n^i\tau_n^j = P_n^j\tau_n^i$ \cite{reif_text, dyk_quantumstats}.

When the lifetimes of the two states are equal, $\tau_n^i = \tau_n^j$, the probability of finding the mode in either state is 1/2. This point is referred to as the kinetic phase transition (KPT) point \cite{dyk_attractors, dyk_hfsr, chan_supernarrow} and is reminiscent of the phase transition point in a thermodynamic system. In the bistable Duffing resonator, the KPT is observed at a specific force detuning which depends only upon the drive power and the physical parameters of the system \cite{dyk_hfsr, dyk_scaling,defoort_scaling, chan_barrier, cleland_switching, vijay_bifamp, qiao_faults}. At the  low noise limit, the KPT and steady-state probabilities are independent of the applied noise \cite{dyk_theory, dyk_escape}. On either side of the KPT, the transition rates are imbalanced and one of the states becomes exponentially dominant. At drive frequencies above the KPT, (i.e. approaching $\Delta \omega_U$) the $\downarrow$ state dominates; at drive frequencies below the KPT point, (i.e. approaching $\Delta \omega_L$) the $\uparrow$ state is dominant \cite{dyk_hfsr}. 

We demonstrate the behavior of the state probabilities in the first mode ($n=1$)  in Fig. \ref{fig:single-switching}c. Here, $P_1^\uparrow$ and $P_1^\downarrow$ are plotted as a function of $\Delta \omega_1^*$ at a constant drive and rms noise amplitude. In the experiments, we measured the  time spent in the $\uparrow$ and $\downarrow$ states at each force detuning, and  calculated the corresponding probabilities by dividing the total time spent in each state by the total duration of the time trace, 120 s. We identify the KPT at the crossing point of the two state probabilities, indicated by the dashed red line. The results confirm that the $\downarrow$ state becomes exponentially dominant above the KPT and the $\uparrow$ state is dominant below. For these specific drive conditions, we find the KPT at a dimensionless detuning of $\Delta \omega_1^* \approx 30.5$.  We confirm that similar observations were made for the $n=4$ mode.

\section{Coupled Nonlinear Nanomechanical Modes}
\label{sec:coupled-modes}
\subsection{Dispersive Mode Coupling}
\label{sec:dispersion}

Having established the behavior of independent modes, we return to the single mode discussion in \S\ref{sec:single-mode} and now consider a beam driven in multiple modes simultaneously. Following standard steps, the  equation of motion for mode $n$ can be obtained as \cite{matheny_modes, dyk_meso, chan_coupled, atakan_coupling},
\begin{eqnarray}
        \label{eq:eom_multi}
    \ddot{x}_n + 2\Gamma_n\dot{x}_n + \left({\Omega_n}^2 + \sum_m \alpha_{nm}{x_m}^2 \right)x_n = \\ \nonumber
     \frac{F_n}{M_n}\cos{\left[\left( \Omega_n + \Delta \omega_n\right) t\right]}.
\end{eqnarray}
Here, the sum is over all modes $m$, including $m=n$.  The assumption is that, for $m\neq n$, the mode is driven at a single frequency with a constant-amplitude sinusoidal and the contribution of undriven modes can be neglected. The last term on the left-hand side of Eq. 13 couples the effective eigenfrequency of mode $n$ to the amplitudes of the other modes ($m\neq n$), a phenomenon known as dispersive coupling. As before, Eq. \ref{eq:eom_multi} can be non-dimensionalized and solved perturbatively using the inverse quality factor $\epsilon = {Q_n}^{-1}$ as the small parameter. Doing so provides an expression for the steady-state amplitude $A_n$ of each mode in terms of   $A_n$ itself and the steady-state amplitudes of the other modes $A_m$ ($m \neq n$).  This expression contains two types of nonlinearity: the \textit{intramodal} cubic nonlinearity ${A_n}^3$ ($m=n$), and a quadratic nonlinearity of the form $A_n{A_m}^2$ for \textit{intermodal} interactions ($m \neq n$) \cite{matheny_modes, atakan_coupling, sader_intermodal}. These expressions are provided in the Appendix.

As discussed above in \S\ref{sec:single-mode}, the \textit{intramodal} term gives rise to the Duffing response where the frequency at maximum amplitude shifts relative to the  eigenfrequency. The \textit{intermodal} effect, however, can be interpreted as a perturbation of the eigenfrequency of mode $n$ itself by the amplitudes of the other modes \cite{matheny_modes, westra_modal}. We write the intermodal perturbation of the eigenfrequency $\Omega_n$ of mode $n$ by the amplitude  $A_k$ of another mode ($k\neq n$) as \cite{matheny_modes,dyk_meso},
\begin{eqnarray}
    \label{eq:freq_shift}
    \Delta \Omega_{nk} &=& {1 \over 4} \frac{\alpha_{nk}}{\Omega_k}{A_k}^2,
\end{eqnarray}
which is nonlinear in  $A_k$ and depends on the intermodal Duffing coefficient $\alpha_{nk}$. An equivalent expression can be written for $\Delta \Omega_{kn}$. The physical implication is that, when $A_k \gg A_n$, the onset of the mode $n$ resonant response is shifted to higher frequencies and \textit{vice versa}. If mode $k$ is driven in the asymptotic limit on the Duffing backbone, where Eq. \ref{eq:amp_approx} applies, we can eliminate $A_k$ from Eq. 14 and  express   $\Delta \Omega_{nk}$ in terms of the drive frequency detuning $\Delta \omega_k$ of mode $k$, yielding
\begin{equation}
    \label{eq:freq_detuning}
    \Delta \Omega_{nk}\approx \frac{2}{3}\frac{\alpha_{nk}}{\alpha_{kk}}\Delta \omega_{k},
\end{equation}
which is linear in the drive frequency detuning of the nonlinear, high-amplitude mode $k$.

\begin{figure}
    \centering
    \includegraphics[]{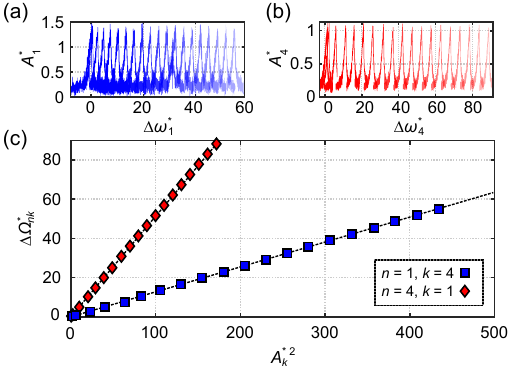}
    \caption{\label{fig:dispersion-lin} (a \& b) Forward frequency sweeps of linear mode 1 and 4 responses at varied amplitudes of the other mode. In (a), mode 1 was driven weakly at -23.5 dBm ($F_1^*=1.4$) while mode 4 was driven strongly at -4.8 dBm ($F_4^*=18.4$), with mode 1 amplitude remaining at $\approx 4$ nm. The frequency step for mode 4 was 6 kHz. In (b), mode 4 was driven weakly at -16.8 dBm ($F_4^*=1.15$) while mode 1 was driven strongly at -7 dBm ($F_1^*=62$), with mode 4 amplitude remaining at $\approx 7$ nm. The frequency step for mode 1 was 10 kHz. (c) The dimensionless eigenfrequency shift $\Delta \Omega_{nk}^*$ relative to $\Omega_n$ in units of linewidth,  where $n,k=1,4$ and $n \neq k$, as a function of the  dimensionless intermodal amplitude squared. Dashed lines show linear fits. }
\end{figure}

Figure \ref{fig:dispersion-lin} shows  the intermodal couplings in our system.  In Fig. \ref{fig:dispersion-lin}a, we measure the frequency response of mode $n=1$ under a weak linear drive, while mode $k=4$ is driven at different amplitudes in its nonlinear regime. In this experiment, we initialize mode $k=4$ by adjusting the drive frequency so that it is driven on its Duffing backbone and its amplitude-frequency dependence is governed by Eq. \ref{eq:amp_approx}. We then increment the mode $k=4$ drive frequency by a set amount. At each mode $k=4$ drive frequency step, we pause and measure the frequency response of mode $n=1$ using a forward frequency sweep and record the position of maximum amplitude relative to the unperturbed eigenfrequency $\Omega_n$. In Fig. \ref{fig:dispersion-lin}a, the  mode 1 frequency response shifts to higher frequencies as the mode $4$ amplitude increases, where the $x$-axis is presented in terms of the dimensionless detuning of the drive. Identical behavior is observed when we repeat this experiment for a strongly-driven mode 1 and weakly-driven mode 4, as shown in Fig. \ref{fig:dispersion-lin}b. When these measurements are properly calibrated, the shift in eigenfrequency $\Delta \Omega_{nk}$ can be obtained as a function of ${A_4}^2$ and ${A_1}^2$, respectively, as shown in the dimensionless plot in Fig. \ref{fig:dispersion-lin}c, where $\Delta \Omega_{nk}^*$ is the dimensionless frequency shift of mode $n$ (in units of $2\Gamma_n$) relative to the unperturbed eigenfrequency $\Omega_n$. The frequency shift is quadratic in the intermodal amplitude, as expected from Eq. \ref{eq:freq_shift}, and provides the intermodal Duffing coefficients of $\alpha_{14}/4\pi^2 = 17.7\times 10^{-5} ~\rm MHz^2/nm^2$ and $\alpha_{41}/4\pi^2 = 17.8 \times 10^{-5} ~\rm MHz^2/nm^2$.  To within experimental precision, $\alpha_{14} \approx \alpha_{41}$  \cite{dyk_meso, vinante_mixing}. 

We now turn to the more interesting question of when both modes are driven nonlinearly. To lowest order, two effects are superposed here. In \S\ref{sec:single-mode} and Fig. \ref{fig:duffing}a, we have established that, when a given mode $n$ is driven on the Duffing backbone, there is a prescribed relationship between the amplitude and the force detuning, i.e., Eq. \ref{eq:amp_approx}. On the other hand, the eigenfrequency of the given mode $n$ will increase when the amplitude of the other mode $k$ increases, as is evident in Fig. \ref{fig:dispersion-lin}, due to the intermodal coupling. This interaction is described by Eq. \ref{eq:freq_shift} and  shifts the onset of the Duffing response of the given mode $n$ to higher frequencies with  increasing amplitude of the other mode $k$. After the onset of the Duffing response of the given mode $n$, the amplitudes of modes $k$ and $n$ become comparable and the interaction is less trivial.

To observe the intramodal and intermodal effects in tandem, we drive both modes strongly at different combinations of drive frequencies, such that both modes trace their respective Duffing backbones. We form a $9 \times 30$ grid of drive frequencies for modes 1 and 4, respectively, with the lowest drive frequency in each mode set below its unperturbed eigenfrequency. To reach each grid point, we fix the drive amplitudes at $F_1 = -10$ dBm ($F_1^*=40$) and $F_4 = -6.7$ dBm ($F_4^*=11.7$) and start below the resonance frequency of each mode, i.e., $\Delta \omega_1, \Delta \omega_4  < 0$. The drive amplitude is set such that both modes are in the nonlinear regime. We then alternate between increasing the drive frequency of each mode quasi-statically so that each mode remains in its $\uparrow$ state as $\Delta \omega_1$ and $\Delta \omega_4$ approach their target values. At the target grid point, the amplitudes of both modes are measured for 1 second. To move on to the next grid point, the drive frequencies are reset to $\Delta \omega_1, \Delta \omega_4  < 0$ and the process is repeated. This process is critical to ensure that  both modes remain in their respective $\uparrow$ states on their Duffing backbones.

\begin{figure*}
    \centering
    \includegraphics[]{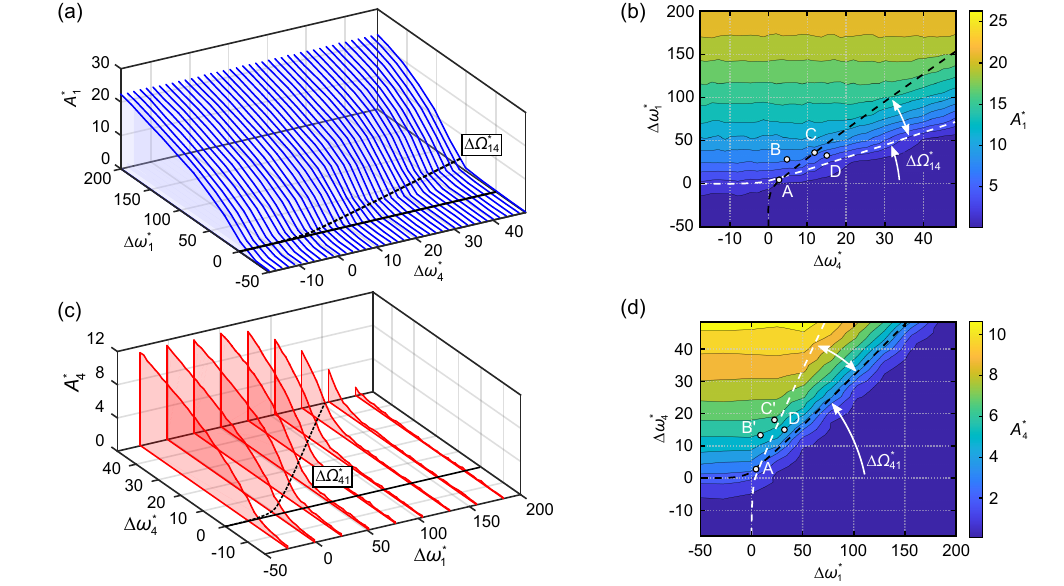}
    \caption{\label{fig:dispersion-nl} (a) Waterfall plot of $A_1^*$ as a function of $\Delta \omega_1^*$ and $\Delta \omega_4^*$. Modes 1 and 4 were driven at -10 dBm ($F_1^*=40$) and -6.7 dBm ($F_4^*=11.7$), respectively, with maximum corresponding amplitudes of $A_1\approx110$ nm and $A_4\approx20$ nm. Each slice represents a constant mode 4 drive frequency. The frequency steps for modes 1 and 4 were 3 kHz and 1.5 kHz, respectively. The unperturbed and perturbed eigenfrequencies of mode $n=1$ are shown as solid and dashed black lines, respectively. (b) Contour map of the data in (a). The perturbed eigenfrequencies of mode $n=1$ and 4 are shown as dashed white and black lines, respectively, calculated as described in the Appendix. The dispersively coupled regime remains in between the two dashed lines marked by a double-sided arrow. Points  A, B, C, D respectively correspond to the monostable, bistable, tristable, and coupled regions presented in Figs. \ref{fig:mono}, 8, 9, and \ref{fig:coupled}. (c) Waterfall plot of $A_4^*$ as a function of the modal drive frequencies, recorded simultaneously with the data in part a. Here, each slice represents a constant mode 1 drive frequency. The unperturbed and perturbed eigenfrequencies of mode $n=4$ are shown as solid and dashed black lines, respectively. (d) Contour map of the data in part c. The $xy$ axes are rotated and flipped relative to the contour map in (b). Here, the perturbed eigenfrequencies of modes $n=1$ and 4 arerespectively shown as dashed white and black lines that are identical to those plotted in (b). Points labeled A, B$'$, C$'$, and D respectively correspond to the monostable, bistable, tristable, and coupled regions presented below in Figs. \ref{fig:mono}, \ref{fig:bi}, \ref{fig:tri} and \ref{fig:coupled}.}
\end{figure*}

The results from the above-described experiment are shown in Fig. \ref{fig:dispersion-nl}. Fig. \ref{fig:dispersion-nl}a-b and Fig. \ref{fig:dispersion-nl}c-d respectively show the dimensionless amplitudes $A_1^*$ and $A_4^*$ at each combination of dimensionless detunings $\Delta \omega_1^*$ and $\Delta \omega_4^*$. Starting with the lower-most corner of Fig. \ref{fig:dispersion-nl}a at the point $(\Delta \omega_1^*,\Delta \omega_4^*)=(-50,-20)$, for instance, mode 1 is observed to be out of resonance. As $\Delta \omega_1^*$ is increased by moving along the left edge, the mode traces the unperturbed  Duffing response, as the force amplitude is large and intermodal contribution from mode 4 is negligible. If $\Delta \omega_4^*$ is increased by moving along the right axis, mode 1 still traces the Duffing curve but the Duffing response starts at a larger  $\Delta \omega_1^*$ value. Similar features are observable in Fig. \ref{fig:dispersion-nl}c for mode 4, when the $\Delta \omega_1^*$-$\Delta \omega_4^*$ plane is rotated by $90^\circ$ and flipped. If the same data are plotted in 2D contour plots (Fig. \ref{fig:dispersion-nl}b and d), we observe that the frequency for the onset of the Duffing response for either mode increases with the drive frequency of the other mode. 

The loci of the perturbed eigenfrequencies of mode 1 and mode 4 are shown by the dashed curves on the plane labeled $\Delta \Omega^*_{14}$ and $\Delta \Omega^*_{41}$. These curves are approximately  the linear shifts  shown above in Fig. \ref{fig:dispersion-lin}. The exact expression for each curve can be obtained by inserting the upper-branch analytical amplitude solution in Eq. \ref{eq:amp_resp} into Eq. \ref{eq:freq_shift}. This captures the smooth transition between the perturbed and unperturbed eigenfrequencies near the origin at $(0,0)$. For the particular drive amplitudes used in Fig. \ref{fig:dispersion-nl}, the intermodal effect persists in the region bounded by the two loci and defines a region where the two modes interact strongly through the dispersive interactions. The mode amplitudes in this region depend strongly on the drive frequencies of both modes. Outside of this region, the intramodal effect on one of the modes dominates, as seen in the upper right corner of the data for mode 1. Here, one mode is weakly perturbed and follows its normal Duffing response while the other mode is strongly perturbed and remains at low amplitude. 

\subsection{Coupled States, Multistability and Inter-state Transition Dynamics}
\label{subsec:multistability}

\begin{figure*}
    \includegraphics[]{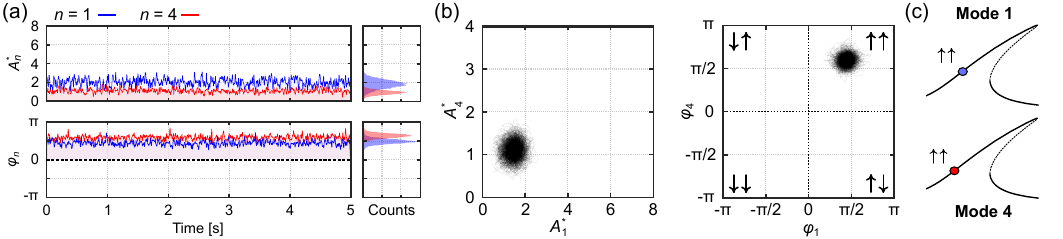}
    \caption{\label{fig:mono} (a) Time traces and histograms recorded at point A in Fig. \ref{fig:dispersion-nl}b with $(\Delta \omega_1^*, \Delta \omega_4^*)=(5,3)$. Here, both modes are driven below their bifurcation frequencies and are monostable. We apply noise power densities of -70 dBm/Hz and -75.2 dBm/Hz to modes 1 and 4 respectively. The oscillatory and rms noise amplitudes for mode 1 are approximately 9 nm and 3.4 nm; for mode 4 they are 1.8 nm and 0.5 nm. (b) Lissajous plots of the mode amplitudes and phases corresponding to the time trace in part a with a total duration of 30 s. (c) Duffing responses for the two modes illustrating the drive conditions in the monostable regime; the steady-state mode amplitudes are indicated with colored dots.
    }
\end{figure*}

As discussed in \S\ref{sec:bistability}, with no intermodal interactions (Fig. \ref{fig:single-switching}), each mode can be treated as a non-equilibrium two-level system, with $P_n^i$ being the probability of finding  mode $n$ in state $i$, where $i =$ $\uparrow$ or $\downarrow$ and $n=1$ or 4. The dispersive coupling between the modes suggests that the state probabilities of a given mode  will depend on the state and drive conditions of the other mode. We therefore treat the two nonlinear nanomechanical modes as a single  system consisting of two interacting two-level subsystems. The coupled system can then reside in one of four available states:  $\uparrow\uparrow$, $\uparrow\downarrow$, $\downarrow\uparrow$, or $\downarrow\downarrow$, where the first and second arrows respectively correspond to the nonlinear branch occupied by mode 1 and 4. In the rest of the paper, we explore the dynamics of transitions between these states. To this end, we repeat the experiments in \S\ref{sec:dispersion} by forming a $41 \times 41$ grid of drive frequencies, which start below resonance for each mode and span the single-mode dominated and dispersively-coupled regimes discussed in \S\ref{sec:dispersion}. We drive each mode with a constant sinusoidal drive, as in Fig. \ref{fig:dispersion-nl}, i.e., $F_1^*=40$ and $F_4^*=11.7$, and add narrowband Gaussian noise to each drive. The maximum amplitude reached by both modes in these experiments is $\sim 60$ nm; the rms noise amplitude is measured to be $\sim 4$ nm and $\sim 0.6$ nm in modes 1 and 4, respectively \cite{myreference}. Relevant experimental parameters are included in the figure captions. 

At each combination of drive frequencies, we record a 30 s time trace of the modal amplitudes and phases. The presence of noise enables the system to undergo noise-induced transitions between its stable states. Accordingly, the system no longer needs to be initialized in a particular state and we simply sweep the drive frequencies to the next grid point after each recording. We discretize each mode signal into $\uparrow$ and $\downarrow$ states at each time step using a phase threshold of $\phi = 0$, where we have shifted the phase by $\pi/2$ so that $\phi > 0$ is assigned $\uparrow$ and $\phi < 0$ is assigned $\downarrow$. The two discretized mode signals are then combined to assign a coupled state $\uparrow\uparrow$, $\uparrow\downarrow$, $\downarrow\uparrow$, or $\downarrow\downarrow$ to the system at each time step. We perform all statistical measurements on this coupled-state signal.

We begin in Fig. \ref{fig:mono}a by presenting a sample time trace at point A in Figs. \ref{fig:dispersion-nl}d and \ref{fig:kpt}a, where $(\Delta \omega_1^*, \Delta \omega_4^*)=(5,3)$. At this force level of $(F_1^*, F_4^*)\approx(40,11.7)$, both modes are below the onset of bistability---which we found to occur at $\Delta \omega_{1L}^* = 8.3$ and $\Delta \omega_{4L}^* = 4.4$ in Fig. \ref{fig:duffing}b---and are monostable on their upper branches. The histograms show that each mode exhibits Gaussian fluctuations about its stable state. We plot the mode amplitudes and phases in Lissajous plots in the $A_1^*$-$A_4^*$ and $\phi_1$-$\phi_4$ planes in Fig. \ref{fig:mono}b, and observe a single stable node in both data plots. In the $\phi_1$-$\phi_4$ plot, we identify the node in the upper right quadrant as the $\uparrow\uparrow$ state, since $\phi_{1}, \phi_4 \approx {\pi \over 2}$. Each individual mode can be visualized on its respective Duffing response curve in Fig. \ref{fig:mono}c by its amplitude at the fixed drive frequencies of $\Delta \omega_1^*$, and $\Delta \omega_4^*$, where both modes are below the onset of bistability. We note that it may not be strictly accurate to assign $\uparrow$ and $\downarrow$ states in the monostable region. However, due to hysteresis, smooth operations (such as frequency sweeps) starting from the monostable region below resonance can only transition to a bistable $\uparrow$ state; operations starting above the bistable region can only transition to the $\downarrow$ state. In addition to the phase difference between the two states, this justifies assigning $\uparrow$ and $\downarrow$ states to the solution branches outside the bistable region.

\begin{figure*}
    \includegraphics[]{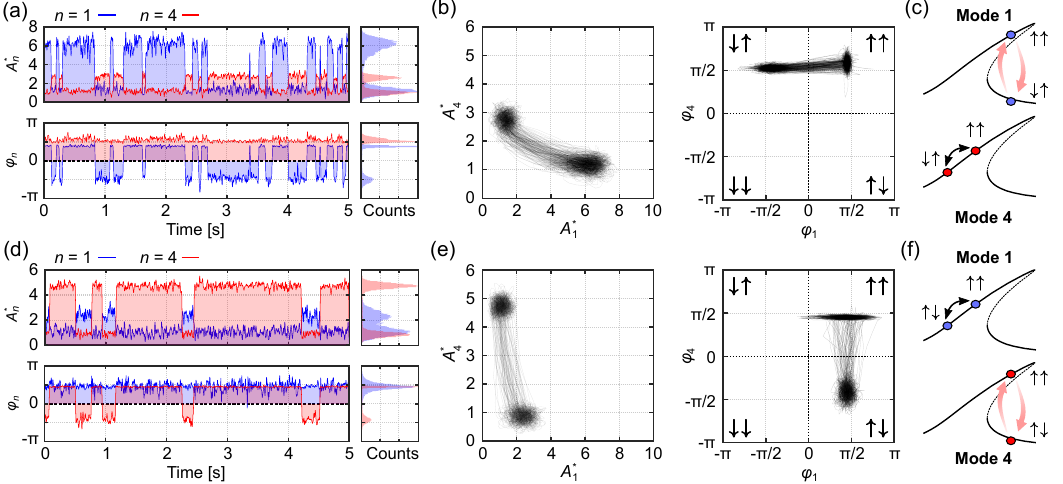}
    \caption{\label{fig:bi} (a \& d) Time traces and histograms recorded at point B \& B' in Fig. \ref{fig:dispersion-nl}b and d respectively, which are the points $(\Delta \omega_1^*, \Delta \omega_4^*)=(28,5)$ and $(\Delta \omega_1^*, \Delta \omega_4^*)=(9.5, 13.5)$. Here, one mode is driven above the onset of bistability and the other mode remains monostable. (b \& e) Lissajous plots of the mode amplitudes and phases in the bistable regime corresponding to the time traces, with a total duration of 30 seconds. (c \& f) Duffing model of the bistable regions for the two modes. The mode amplitudes at fixed drive frequency are indicated with colored dots. The two stable states of the bistable mode perturb the eigenfrequency of the monostable mode relative to a fixed drive frequency and lead to two different steady-state amplitudes for the same nonlinear state. For comparison, the maximum oscillatory and rms noise amplitudes are approximately 30 nm and 3.4 nm for mode 1, and 5 nm and 0.5 nm for mode 4, respectively in (a). In (d), these values are respectively 12 nm and 3.4 nm for mode 1, and 9 nm and 0.5 nm for mode 4.
    }
\end{figure*}

Starting from the monostable region, we can drive one of the modes into bistability by increasing its drive frequency. Figs. \ref{fig:bi}a-c present the dynamics at point B in Figs. \ref{fig:dispersion-nl}b \& \ref{fig:kpt}e, where $(\Delta \omega_1^*, \Delta \omega_4^*)=(28,5)$. Here, mode 1 is above its onset of bistability and can access both the $\uparrow$ and $\downarrow$ states; mode 4 is below the onset of bistability and remains in its $\uparrow$ state. The coupled system is therefore able to access the $\uparrow\uparrow$ and $\downarrow\uparrow$  states and is \textit{bistable}. The time-trace in Fig. \ref{fig:bi}a shows that mode 1 undergoes transitions between its $\uparrow$ and $\downarrow$ states that manifest as two distinct levels in both amplitude and phase. Mode 4 undergoes simultaneous changes in amplitude in response to the change in tension as mode 1 jumps between its $\uparrow$ and $\downarrow$ states, but does not change its phase. These unique features can also be observed when the data are plotted in  Lissajous plots \cite{pikovsky_synch} in the $A_1^*$-$A_4^*$ and $\phi_1$-$\phi_4$ planes, shown in Fig. \ref{fig:bi}b. The $\phi_1$-$\phi_4$ plot shows that the system remains in the upper two quadrants, which are identified as the $\uparrow\uparrow$ and $\downarrow\uparrow$ states. In the $A_1^*$-$A_4^*$ plane, rapid transitions occur between the two nodes along a single curved trajectory that indicates correlated dynamics of the two modes as they respond to the changing instantaneous tension in the beam. The dispersive interaction is illustrated by the two Duffing curves in Fig. \ref{fig:mono}c. Mode 1 is driven above its bifurcation and can access both branches; the different amplitudes of these two states affect the eigenfrequency of mode 4 and lead to two distinct drive detunings and steady-state amplitudes relative to the fixed drive frequency. The change in mode 4 amplitude has a small reciprocal effect on mode 1 that slightly perturbs its amplitude in the $\downarrow\uparrow$ state. These interactions suggest that the proper model of the dynamics is from the perspective of the coupled system, which transitions between two states that are associated with particular amplitudes and phases of both modes, rather than two independent modes interacting. A very similar behavior is observed when mode 4 is driven to bistability, shown in Figs. \ref{fig:bi}d-f, corresponding to point B$'$ in Figs. \ref{fig:dispersion-nl}b \& \ref{fig:kpt}f, where $(\Delta \omega_1^*, \Delta \omega_4^*)=(9.5, 13.5)$.  The two states now lie in the right-hand quadrants in the $\phi_1$-$\phi_4$ plane (Figs. \ref{fig:bi}e) and the transitions occur vertically in the $A_1^*$-$A_4^*$ plane (Figs. \ref{fig:bi}e). 

\begin{figure*}
    \includegraphics[]{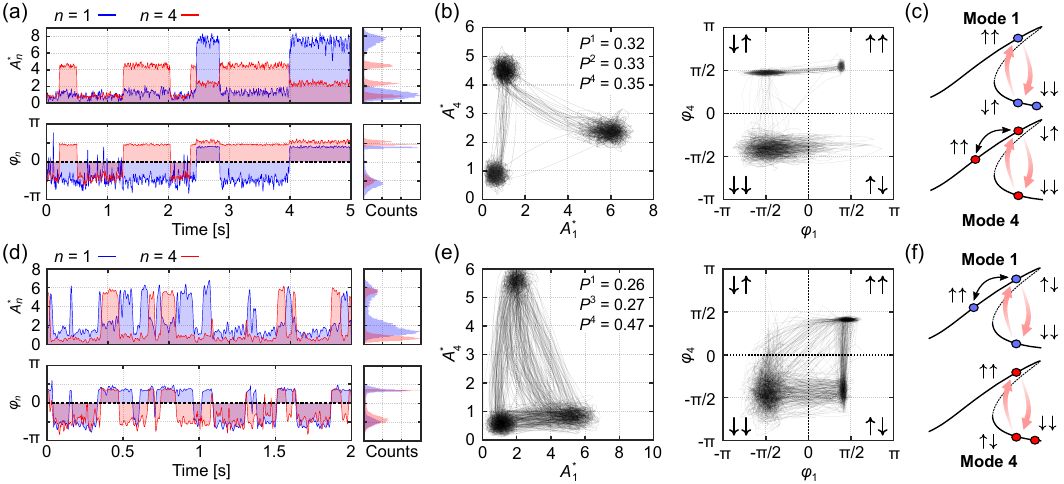}
    \caption{\label{fig:tri} (a \& d) Time traces and histograms recorded at point C and C$'$ in Figs. \ref{fig:dispersion-nl}b \& d, respectively, with coordinates $(\Delta \omega_1^*, \Delta \omega_4^*)=(36, 12)$ and $(\Delta \omega_1^*, \Delta \omega_4^*)=(23, 18)$. These points also correspond to the points C and C$'$ in Fig. \ref{fig:kpt}h and are the mode 1 and mode 4 triple points, where the probability of finding the system in each state is roughly equal. (b \& e) Lissajous plots of the mode amplitudes and phases at the mode triple points in a total duration of 30 s. (c \& f) Duffing models of tristability. Each state corresponds to a different position on the Duffing response curves relative to the fixed drives.  This leads to three distinct amplitudes for the system at each triple point.  The maximum oscillatory and rms noise amplitudes are approximately 37 nm and 3.4 nm for mode 1, and 9 nm and 0.5 nm for mode 4, respectively, in (a). In (d), these values are respectively 27 nm and 3.4 nm for mode 1, and 10 nm and 0.5 nm for mode 4.
    }
\end{figure*}

Proceeding  from the mode 1 bistable region (Figs. \ref{fig:bi}a-c), we now increase the drive frequency of the monostable mode ($n=4$) until it approaches its  onset of bistability. Now, the tension in the beam due to the jumps of the already bistable mode ($n=1$) can modulate the monostable mode ($n=4$) into and out of bistability. These coupled transitions then allow the system to access three distinct system states: $\uparrow\uparrow$, $\downarrow\downarrow$, and  $\downarrow\uparrow$, as shown in Fig. \ref{fig:tri}a-c. This behavior is shown in the time trace in Fig. \ref{fig:tri}a, which was recorded at point C in Figs. \ref{fig:dispersion-nl}b \& \ref{fig:kpt}h, at coordinates  $(\Delta \omega_1^*, \Delta \omega_4^*)=(36, 12)$. As we discuss further below, at this point, the probabilities of the three states are equal, and we refer to this point as the \textit{mode 1 triple point}. The time trace shows that both mode 1 and mode 4 undergo changes in amplitude and phase---unlike in the bistable regime---and we  identify three system states, $\uparrow\uparrow$, $\downarrow\uparrow$, and $\downarrow\downarrow$. These states appear as three peaks in the mode 4 amplitude histogram and form three distinct nodes in the corresponding $A^*_1$-$A_4^*$ and $\phi_1$-$\phi_4$ planes, as shown in Fig. \ref{fig:tri}b. The system transitions directly between these nodes along three separate curved pathways. For this particular set of drive conditions we see very few $\downarrow\downarrow$ to $\uparrow\uparrow$ transitions; the overall probability of each state, however, is nearly equal.  A model of the dynamics in the vicinity of the mode 1 triple point is presented in Fig. \ref{fig:tri}c. Starting in the $\uparrow\uparrow$ state, we observe that mode 1 is bistable while mode 4 is monostable. When mode 1 transitions into its $\downarrow$ state, the coupled system transitions to the $\downarrow\uparrow$ coupled state. The reduced tension of the $\downarrow$ state in mode 1 lowers the mode 4 eigenfrequency such that mode 4 is now in the bistable regime and is able to undergo a transition to its lower state as well. If mode 4 transitions to its lower branch, the coupled system enters the $\downarrow\downarrow$ state. From the $\downarrow\downarrow$ state, both modes can undergo a transition back to their upper branches, as we observe in the Lissajous plots in Fig. \ref{fig:tri}b. The bistability of mode 4 is thus modulated by the occupied state of mode 1. Identical dynamics occur at the mode 4 triple point, which is point C$'$, $(\Delta \omega_1^*, \Delta \omega_4^*)=(23, 18)$, in Figs. \ref{fig:dispersion-nl}d \& \ref{fig:kpt}h. These results are shown in Figs. \ref{fig:tri}d-f, where we see the $\uparrow\downarrow$ state instead of the $\downarrow\uparrow$ state.

\begin{figure*}
    \includegraphics[]{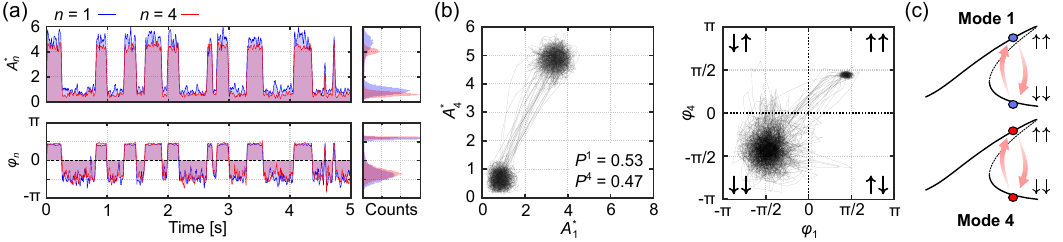}
    \caption{\label{fig:coupled} (a) Time traces and histograms recorded at point D in Fig. \ref{fig:dispersion-nl} with $(\Delta \omega_1^*, \Delta \omega_4^*)=(32, 15)$. Here, the two modes undergo coupled transitions to and from their upper and lower branches. (b) Lissajous plots of the mode amplitudes and phases. (c) Duffing responses for the two modes illustrating the drive conditions in the coupled bistable regime, where the steady-state mode amplitudes are indicated with colored dots. Both modes are driven above their bifurcation frequency and are bistable. The maximum oscillatory and rms noise amplitudes are approximately 25 nm and 3.4 nm for mode 1, and 7 nm and 0.5 nm for mode 4, respectively.
    }
\end{figure*}

From the mode 1 triple point, if we continue to increase the drive frequency of (formerly-monostable) mode 4, we enter the dispersively-coupled region discussed in \S\ref{sec:dispersion}. Within this region, for instance point D in Figs. \ref{fig:dispersion-nl}b \& \ref{fig:kpt}g at $(\Delta \omega_1^*, \Delta \omega_4^*)=(32, 15)$, we observe markedly different behavior that we present in Fig. \ref{fig:coupled}. In the time-trace, we see that the modes undergo transitions between their upper and lower states in a highly correlated manner, with no asymmetric states, $\uparrow\downarrow$ and $\downarrow\uparrow$, observed. The $A_1^*$-$A_4^*$ and $\phi_1$-$\phi_4$ plots show that the system transitions directly between two nodes that correspond to the $\uparrow\uparrow$ and $\downarrow\downarrow$ states. The transition pathways show no indication of an intermediate step within the time resolution of the lock-in amplifier. The dynamics in this region are not easily captured by the perturbed Duffing plots used previously due to the large intermodal perturbation of both modes, which further demonstrates the breakdown of the independent-mode picture. We speculate that the large perturbation of the modes in the asymmetric states may make them unstable. We further suggest that there may exist two regimes of behavior: a directly correlated regime, as we present here, where the asymmetric modes are fully unstable and cannot be accessed; and a statistically-correlated regime, where the lifetime of the asymmetric modes becomes very short due to modification of the activation barriers---similar to the phenomenon of stochastic resonance \cite{nicolis_dualSR, buks_amp, dyk_escape}. These regimes may be distinguishable at lower noise powers and are beyond the scope of this paper.

\subsection{State Probabilities and KPT Curves}
\label{sec:kpt}

In order to further discuss the physics of coupled mode transitions, we return to the single-mode case shown in Fig. \ref{fig:single-switching}. We confirmed in \S\ref{sec:bistability} that the probability of observing mode $n$ in state $i$, $P_n^i$, is a function of the detuning of the modal drive, $\Delta \omega_n^*$. When two modes are coupled dispersively, we have shown above in \S\ref{subsec:multistability} that four possible states, $\uparrow\uparrow$, $\downarrow\uparrow$, $\uparrow\downarrow$, and $\downarrow\downarrow$, emerge, which  we enumerate as $r = 1,2,3,4$, respectively. The coupled states have probabilities ${\cal P}^r$, which should now depend on both $\Delta \omega_1^*$ and $\Delta \omega_4^*$ due to the dispersive interactions presented in  \S\ref{sec:dispersion}. The coupled-mode probability plots corresponding to Fig. \ref{fig:single-switching}c are thus two-dimensional in drive frequencies of the two modes involved. 

\begin{figure*}
    \includegraphics[]{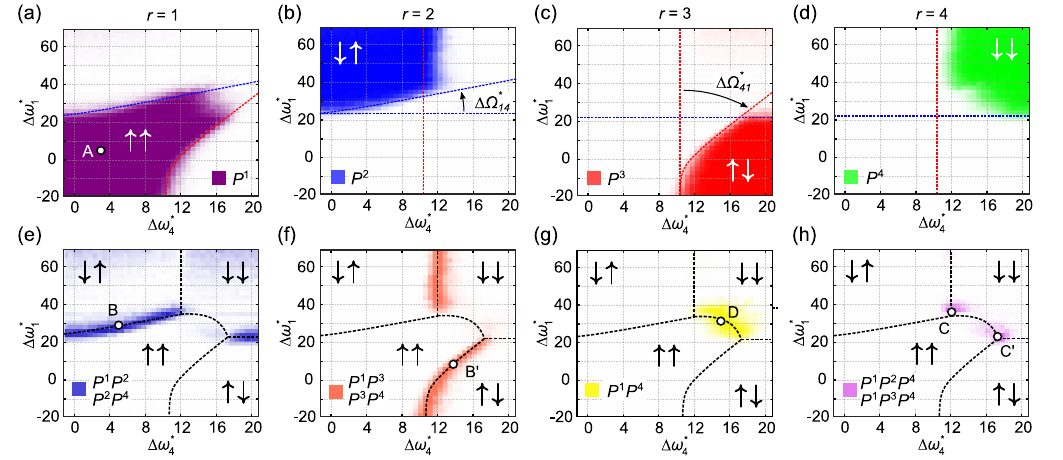}
    \caption{\label{fig:kpt} (a-d) Probability of each coupled state  $r =1$ to 4 in the $\Delta \omega_1^*$-$\Delta \omega_4^*$ plane. The system  occupies the state $r$ with a probability of 1 within the solid colored region  and 0 in the transparent region.  The blue and red lines indicate the perturbed or unperturbed KPT curves for the single-mode bistable transitions. (e-h) Maps of the joint probabilities corresponding to the indicated transitions. Solid color represents maximum joint probability, and white indicates a joint probability of zero. In this experiment, the modes were driven sinusoidally with powers of $-10$ dBm ($F_1^*=40$) and $-6.7$ dBm ($F_4^* =11.7$). The noise power density was -68 dBm/Hz for mode 1 and -74.3 dBm/Hz for mode 4, corresponding to rms displacements of 4 nm and 0.6 nm respectively. The five KPT curves are plotted as black dashed lines. }
\end{figure*}

The probability of each coupled state $r$ is determined by dividing the total time spent in the state by the length of the trace, 30 s. For example, in Fig. \ref{fig:bi}, the probability of the $\uparrow\uparrow$ state is approximated by determining the total time spent in the upper right quadrant in the $\phi_1-\phi_4$ plane and dividing by 30 s. The probabilities of the coupled states, $r = 1,2,3$ and $4$, as a function of $\Delta \omega_1^*$ and $\Delta \omega_4^*$ are shown in Fig. \ref{fig:kpt}a-d, respectively. Solid colored regions correspond to  ${\cal P}^r = 1$ whereas transparent regions correspond to ${\cal P}^r = 0$. The state probabilities form four distinct regions in the $\Delta \omega_1^*$-$\Delta \omega_4^*$ plane; in each region, one of the four states is stable. The monostable time trace shown in Fig. \ref{fig:mono}, for instance, is taken from within the $\uparrow\uparrow$ region, at point A in Fig. \ref{fig:kpt}a. Similar dynamics are observed in any of the four stable quadrants, which differ in the quadrant occupied in the $\phi_1$-$\phi_4$ plane.

To understand the behavior near the boundaries of these stable regions, we again refer to Fig. \ref{fig:single-switching}c and consider the joint probability of the two states, $P_1^\uparrow  P_1^\downarrow$. Away from the KPT point, $P_1^\uparrow  P_1^\downarrow \approx 0$ and the mode is stable in one of the two states. As the KPT point is approached, transitions become appreciable and $P_1^\uparrow  P_1^\downarrow > 0$. At the KPT point, both  $P_1^\uparrow  \approx {1\over 2}$ and $P_1^\downarrow \approx {1 \over 2}$, and the joint probability attains a maximum value of $P_1^\uparrow  P_1^\downarrow \approx {1 \over 4}$.  We generalize this observation and calculate the relevant joint probabilities in the  $\Delta \omega_1^*$-$\Delta \omega_4^*$ plane, with the expectations that KPT curves will be defined where the joint probabilities are maximum and the joint probabilities will approach zero away from these curves. For instance, where two states $r$ and $s$ are contiguous in the plane, we calculate ${\cal P}^r  {\cal P}^s$; in the case of three contiguous states $r$, $s$, and $t$, we calculate ${\cal P}^r  {\cal P}^s  {\cal P}^t$; and so on. The results of these calculations are compiled in Fig. \ref{fig:kpt}e-h. 

In this four-state system, five KPT curves emerge, as shown by the dashed curves in Fig. \ref{fig:kpt}e-h, which form the boundaries between pairs of the stable regions. Where only two regions $r$ and $s$ come into contact, as in Fig. \ref{fig:kpt}e, f, and g, the joint probability  ${\cal P}^r {\cal P}^s$ becomes nonzero as one approaches the KPT curve between the regions. In Figs. \ref{fig:kpt}e, f and g, we plot the joint probabilities of $r = 1$ and $s=2$, $r = 1$ and $s=3$, and $r = 1$ and $s=4$, respectively. Along the nearly-horizontal KPT line at the left edge in Fig. \ref{fig:kpt}e,  transitions between coupled states $\uparrow\uparrow$ and $\downarrow\uparrow$ take place. The dynamics in this region are shown in the mode 1 bistable data presented in Figs. \ref{fig:bi}a-c. The KPT line is slightly tilted along $\Delta \omega_4^*$ (dashed blue lines in Figs. \ref{fig:kpt}a and b), which is well-described by substituting the upper branch amplitude solution to Eq. \ref{eq:amp_resp} into the intermodal frequency shift in Eq. \ref{eq:freq_detuning}. Similarly, we find a boundary between $r=1$ and 3 along the bottom edge in Fig. \ref{fig:kpt}f,  which involves the state transitions between $\uparrow\uparrow$ and $\uparrow\downarrow$. Again, this boundary is well-described by the intermodal shift, which we plot as  dashed red lines in Figs. \ref{fig:kpt}a and c, and the dynamics are that of the mode 4 bistable case presented in Fig. \ref{fig:bi}d-f.

Along the right and top edges in Figs. \ref{fig:kpt}e and f, respectively, we find KPT points between the $r = 4$ state and the two asymmetric $s = 2$ and $s=3$ states. At these boundaries, we encounter the $\downarrow\downarrow$-$\uparrow\downarrow$ and $\downarrow\downarrow$-$\downarrow\uparrow$ transitions. The intermodal shift in this case is still described by Eq. \ref{eq:freq_detuning} but now uses the lower branch solution to Eq. \ref{eq:amp_resp}, which is much lower in amplitude. The small intermodal shift means that the KPT curves in this case nearly coincide with the KPT points identified along the left and bottom edges, where the mode is on the upper branch but at low amplitude as the drive frequency is below resonance. The unperturbed KPT curves are plotted as straight blue and red dotted lines in Figs. \ref{fig:kpt}b-d. Finally, we find a fifth KPT curve that lies between the $r = 1$ and 4 regions, which we show in Fig. \ref{fig:kpt}h. This curve, which is drawn as a guide to the eye roughly following the observed KPT points, crosses the dispersively coupled region and does not correspond to either of the single-mode KPT curves. Along this curve, we observe the dispersively coupled mode dynamics presented in Fig. \ref{fig:coupled}. 

Three of the five KPT curves intersect at two the points C and C$'$ in Fig. \ref{fig:kpt}h. In the vicinity of these two points, there are regions of tristability, as presented in Fig. \ref{fig:tri}. We quantify these regions by finding the joint probability of the three states. At the intersection points, the probabilities of all three states are approximately equal; in analogy with thermodynamic phase diagrams we refer to these points as \textit{triple points}. At the mode 1 triple point (point C) the $r = 1,2$ and 4 states exist in equilibrium; at the mode 4 triple point (point C$'$) the $r = 1,3$ and 4 states are in equilibrium. The mode 1 triple point does not precisely align with the intersection point, possibly due to a small experimental error.

The state probabilities vary predictably relative to these KPT curves and triple points. Along the KPT curves the two contiguous states remain at equilibrium, with equal probability, even as the relative amplitudes change. Away from the KPT lines, one state becomes exponentially dominant. Finally, around each triple point, the probability of each state varies continuously, such that one of the three states dominates in each adjacent region and different pairs of states dominate along each of the three KPT curves. These results are compiled in the Appendix.

\section{Conclusions}

As we have shown, dispersive interactions between  nonlinear bistable modes can lead to complex dynamics. We experimentally investigated  the dynamics of two nonlinearly-driven  modes of a doubly-clamped nanomechanical beam, with the relevant parameter space being defined solely by  the drive frequencies of the two modes. The dispersive interactions, in turn, were mediated by the tension in the beam.  The interactions culminated in splitting  the parameter space into well-defined regions with significantly different dynamics:   in certain regions, the dynamics was dominated by one of the modes; in others, highly correlated dynamics involving both modes emerged. We observed similarities between this complex nonequilibrium system and an equilibrium thermodynamic system going through a phase transition. Most notably, a triple point emerged, reminiscent of three-phase equilibria, where three states had equal probabilities of occupation. We also observed ``locked jumps" between the two modes, where the modes pulled each other strongly. Based on our experimental results, it will be possible to extend the theoretical KPT description of more complex coupled nonlinear systems. 

More specifically, the multimode dynamics may be understood in light of the Ising model \cite{chan_ising, mahboob_ising} by treating the dynamical $\uparrow$ and $\downarrow$ states of each mode $n$ as spin states and the detuning $\Delta\omega_n$ of the force as the local magnetic field strength on subsystem $n$. In the single-mode case, the transition from the $\uparrow$ to $\downarrow$ state as the external field is strengthened (Fig. \ref{fig:single-switching}c) indicates that the $\uparrow$ state is energetically favored and has an energy that can be related to the KPT frequency. The activation energy of the $\uparrow$ to $\downarrow$ transition depends on the applied force and determines the bifurcation and KPT frequencies by deepening the well in the effective potential (Fig. \ref{fig:duffing}b) \cite{dyk_escape}. The spin-spin coupling, which occurs through the intermodal dispersion, then adds an additional interaction energy component that was found to favor the symmetric $\uparrow\uparrow$ and $\downarrow\downarrow$ states seen in Fig. \ref{fig:kpt}. Thus, the spin coupling constant for the modes examined can be considered to be positive, and the presence of asymmetric states ($\uparrow\downarrow$ and $\downarrow\uparrow$) at high field magnitudes ($\Delta\omega_1^*\gg 1$ or $\Delta\omega_4^*\gg1 $) indicates that the local magnetic field is independent for each mode \cite{chan_ising, mahboob_ising}. Any change in the resonant frequency by altering the tension, mass, or other physical properties can tune the coupling strength. Combined with the ability to tune the activation energy for each mode by varying the force, this system may prove useful for investigating and implementing Ising machines. 

Our experiments may additionally stimulate future work in other directions. The dispersively-coupled dynamics investigated here represents a distinct form of mode coupling compared to typical mechanisms such as internal resonance, which only occurs  under specific conditions when the modes are commensurate \cite{dario_stabilization}. The relatively wide operating frequency range and relaxed mode requirements may make dispersive mode coupling attractive for applications in devices that exhibit a spread in parameters, like eigenfrequency, due to fabrication. While this work was conducted using NEMS resonators, the behaviors observed are expected to be universal for driven multimode systems with a cubic nonlinearity; this includes superconducting resonators, Kerr optical cavities, and analog electrical circuits. In terms of applied research, this work may open up new directions in precision sensing using multi-mode stochastic resonance; mechanical multiplexing and de-multiplexing using coupled transitions; and parametrically-protected mechanical logic in the tristable regime, where the ability of a mode to undergo a transition is controlled by the dynamical state of a second mode. Many points of fundamental significance also remain unexplored, including the variation in transition rates with respect to the noise intensity within one or both modes; spectral properties of the modes undergoing transitions; and a comparison of the sensitivity of single mode and multi-mode systems to external modulations.  

\begin{acknowledgments}
We acknowledge support from the US NSF (Nos. CMMI-2001403, CMMI-1934271, and CMMI-2337507).
We thank M. Ma for the FEM  simulations and H. B. Chan for helpful discussions.
\end{acknowledgments}

\appendix

\section{Duffing Amplitude Response}

Here, we quote the first-order solution to Eq. \ref{eq:eom_multi}. If higher-order correction terms and harmonic responses at multiples of the oscillation frequency are neglected, it can be shown that the response measured in a narrow frequency band around mode $n$ can be expressed as \cite{lc_nems, ochs_thesis}
\begin{equation}
        \label{eq:app_amp_resp}
    {A_n}^2 =   \frac{{{F_n}^2/4{\Omega_n}^2{M_n}^2}}{ \left(\Delta \omega_n - {3\alpha_{nn} {A_n}^2 \over 8\Omega_n} - \sum_m {\alpha_{nm}{A_m}^2 \over 4\Omega_n}\right)^2 + {\Gamma_n}^2}.
\end{equation}
Here, terms with ${\Delta \omega_n}^2$ and $\Gamma_n\Delta \omega_n$ have been neglected and intermodal amplitude contributions have been kept. 

Equation \ref{eq:app_amp_resp} can be rearranged in terms of the force detuning,
\begin{eqnarray}
        \label{eq:app_detuning}
     \Delta \omega_n = {3 \over 8}{\alpha_{nn} \over \Omega_n}{A_n}^2 + \sum_m {1 \over 4}{\alpha_{nm} \over \Omega_n}{A_m}^2 \\ \nonumber \pm \sqrt{{{F_n}^2 \over 4{\Omega_n}^2{M_n}^2{A_n}^2} - {\Gamma_n}^2}.
\end{eqnarray}
At the limit of small dissipation and  $A_n \gg A_m$, we obtain the asymptotic relation
\begin{equation}
    \label{eq:app_aymptote}
    \Delta \omega_n = {3 \over 8}{\alpha_{nn} \over \Omega_n}{A_n}^2,
\end{equation}
which describes the \textit{intramodal} perturbation of the resonant frequency $\Omega_n$ and defines the Duffing backbone. 

To find an approximation for the intermodal perturbation to the eigenfrequency $\Omega_n$ by mode $k$, we consider the  condition when mode $n$ is driven linearly by a force at $\omega_n$ and only mode $k$ is excited out of the sum over $m$ in Eq. \ref{eq:eom_multi}. We re-write the solution in terms of $\omega_n$ and $\Omega_n$ (instead of $\Delta \omega_n$) and obtain
\begin{equation}
    \label{eq:app_amp_resp_2}
    {A_n}^2 = \frac{{F_n}^2}{{4{{{M_n}^2}{\Omega_n}^2}}\left[\left( \omega_n-\Omega_n - {1 \over 4}{\alpha_{nk} \over \Omega_n}{A_k}^2\right)^2 + {\Gamma_n}^2\right]}.
\end{equation}
The  peak occurs when the term in the parentheses in the denominator is zero, which occurs at a drive $\omega_n=\Omega_n+\Delta \Omega_{nk}$. Thus, we find an approximation for the intermodal frequency shift as
\begin{equation}
    \label{eq:app_intermodal}
    \Delta \Omega_{nk} \approx {1 \over 4}{\alpha_{nk} \over \Omega_n}{A_{k}}^2.
\end{equation}


When mode $k$ is driven on its backbone, we can use the asymptotic relation Eq. \ref{eq:app_aymptote} to express the amplitude in terms of the force detuning and the intermodal frequency shift of mode $n$ becomes
\begin{equation}
    \label{eq:app_intermodal_detuning}
    \Delta \Omega_{nk} = {2 \over 3}{\alpha_{nk} \over \alpha_{kk}}{\Omega_k \over \Omega_n}\Delta \omega_k,
\end{equation}
which is a linear function of $\Delta \omega_k$. If mode $k$ is not driven on its backbone, we instead substitute the full analytical expression for $A_k$ along its upper branch which is obtained by directly solving Eq. \ref{eq:app_amp_resp}. Further details are provided in \S \ref{sec:app-kpt}.

\section{Non-dimensionalization\label{app:nd}}

It is convenient to non-dimensionalize Equation \ref{eq:eom_final} when comparing the behavior of multiple modes. We  follow  Lifshitz \& Cross \cite{lc_nems} and make the substitutions for  mode $n$,
\begin{eqnarray}
    \label{eq:app_nondim}
   t_n^* &=& \Omega_nt, \\
   x^*_n &=& x_n\frac{\sqrt{\alpha_{nn}Q_n}}{\Omega_n}, \\  
   F^*_n &=& {F_n\sqrt{\alpha_{nn}{Q_n}^3}\over {M_n\Omega_n}^3},\\
   \Delta \omega^*_n &=& \frac{\Delta \omega_n}{2\Gamma_n} = {Q_n \over \Omega_n}\Delta \omega_n,  
\end{eqnarray}
the physical meanings of which are described in the main text.  We note that there are minor deviations from Lifshitz \& Cross due to differences in the appearance of mass in the equation of motion \ref{eq:eom_final}.  Defining the small parameter $\epsilon = {Q_n}^{-1}$ and substituting in Eq. \ref{eq:eom_final} gives a non-dimensional Duffing equation for a single mode in the form
\begin{equation}
    \label{eq:app_eom_secular}
    {d^2 x_n^* \over d{t^*}^2} + \epsilon{dx_n^* \over dt^*} + x_n^* + \epsilon {x_n^*}^3 = \epsilon F_n^*\cos{\left[(1+\epsilon \Delta \omega^*_n )t_n^*\right]}.
\end{equation}
The first-order amplitude solutions in dimensionless variables are given by
\begin{equation}
    \label{eq:app_amp_resp_nd}
    {A_n^*}^2\left[ \left(2\Delta \omega^*_n - \frac{3}{4}{A_n^*}^2\right)^2 + 1\right] = {F_n^*}^2
\end{equation}
and
\begin{equation}
    \label{eq:app_detuning_nd}
    \Delta \omega_n^* = \frac{3}{8}{A_n^*}^2 \pm {1 \over 2}\sqrt{{{F_n^*}^2 \over {A_n^*}^2} - 1},
\end{equation}
which correspond to Eqs. \ref{eq:app_amp_resp} and \ref{eq:app_detuning} above for a single driven mode.

Similarly, the intermodal frequency shift due to mode $k$ can be expressed in dimensionless form as
\begin{equation}
    \Delta \Omega_{nk}^* = {1 \over 4}\rho_{nk}{A_k^*}^2 = {2 \over 3}\rho_{nk}{\Delta \omega_n^*}
\end{equation}
where $\rho_{nk}$ is a dimensionless unit conversion constant defined by
\begin{equation}
    \rho_{nk} = {\alpha_{nk} \over \alpha_{kk}}{{\Omega_k}^2 \over {\Omega_n}^2}{Q_n \over Q_k}.
\end{equation}

\section{Single-Mode Bistability}
\label{sec:app-bif}

The region of bistability corresponds to the range of parameters where the cubic amplitude equation \ref{eq:app_amp_resp} has three real roots; this occurs when the discriminant of the cubic equation is positive. We set the discriminant equal to zero and solve for the drive force. This gives two solutions for the dimensionless drive force $F^*_n$ squared as a function of the force detuning $\Delta \omega_n^*$,
\begin{equation}
        \label{eq:app_discrim}
    {F^*_{n}}^2 = \frac{16}{9}\Delta \omega^*_{n} + \frac{64}{81}{\Delta \omega^*_{n}}^3 \pm \frac{8}{81}\left( 4{\Delta \omega^*_{n}}^2 - 3\right)^{3/2},
\end{equation}
where the positive and negative solutions correspond to the upper and lower bifurcations, respectively. These two solutions converge at a critical force amplitude of $F^*_c = (4/3)^{5/4}$ and detuning of $\Delta \omega^*_c = \sqrt{3}/2$. Equation \ref{eq:app_discrim} can be simplified at large detuning to give the upper $(\Delta \omega^*_{nU})$ and lower $(\Delta \omega^*_{nL})$ bifurcation points in terms of the drive force,
\begin{equation}
            \label{eq:app_bif_solved}
    \Delta \omega^*_{nU} = \frac{3}{8}{F_n^*}^2 \hspace{1cm} \Delta \omega^*_{nL} = \sqrt[3]{\frac{81}{128}}{F_n^*}^{2/3}.
\end{equation}

\section{Analytical Calculation of the Intermodal Frequency Shift}
\label{sec:app-kpt}
To calculate the exact eigenfrequency perturbation $\Delta \Omega_{nk}$ of mode $n$ due to the amplitude of mode $k$, as shown in Figs. \ref{fig:dispersion-nl} and \ref{fig:kpt}, we seek an analytical expression for $A_k$ to insert into Eq. \ref{eq:freq_detuning}. Since the amplitude of mode $n$ is small when driven below resonance, for $\Delta \omega_n^* < \Delta \Omega_{nk}^*$ we can neglect the reciprocal effect on mode $k$; under these conditions the equation of motion for mode $k$ is independent of mode $n$ while mode $n$ experiences a shift in its effective eigenfrequency. The amplitude solution for mode $k$ is then identical to the single-mode solution in Eq. \ref{eq:amp_resp}.

It is most convenient, however, to begin with Eq. \ref{eq:app_amp_resp_nd} expressed in terms of the perturbing mode $k$. On expansion, we arrive at a cubic equation in ${A_k^*}^2$,
\begin{equation}
{9\over16}\left({A_k^*}^2\right)^3 -3\Delta \omega_k^*\left({A_k^*}^2\right)^2 +\left({4\Delta \omega_k^*}^2 + 1\right)\left({A_k^*}^2\right) - {F_k^*}^2 = 0.
\end{equation}
Using MATLAB, we obtain three solutions for ${A_k^*}^2$ that we write as ${A_k^*}^2={\cal{F}}^{\uparrow}\left(\Delta \omega_k^*,F^*_k\right)$, ${\cal{F}}^{\downarrow}\left(\Delta \omega_k^*,F^*_k\right)$, and ${\cal{F}}^{U}\left(\Delta \omega_k^*,F^*_k\right)$, which represent the upper, lower, and unstable branch solutions of the Duffing equation respectively. 

The dimensionless amplitude on the upper branch for mode $k$ is calculated from the square root of ${\cal{F}}^{\uparrow}$ within the selected range of $\Delta \omega_k^*$ for a given $F_k^*$. For the curves in Figs. \ref{fig:dispersion-nl} and \ref{fig:kpt}, we used the dimensionless forces $F_1^* = 30.3$ and $F_4^* = 11.8$. Next, we calculate the expected perturbation by plugging in the theoretical $A_k^*$ value at each $\Delta \omega_k^*$ into Eq. \ref{eq:app_detuning_nd} to obtain $\Delta \Omega_{nk}^*$ as a function of $\Delta \omega_k^*$. Finally, we calculate the perturbation of frequencies of interest---including the eigenfrequency, KPT point, and lower bifurcation frequency---by adding the frequency of interest to perturbation curve as a constant offset. For the eigenfrequency, which is constant at $\Delta \omega_n^* = 0$, this simply results in a plot of $\Delta \Omega_{nk}$; for the mode 1 KPT point at $\Delta \omega_1^* = 30.5$, this produces a perturbed KPT curve that begins at $\Delta \omega_1^* = 30.5$.

\section{Trends in Transition Dynamics}

\begin{figure*}
    \includegraphics[]{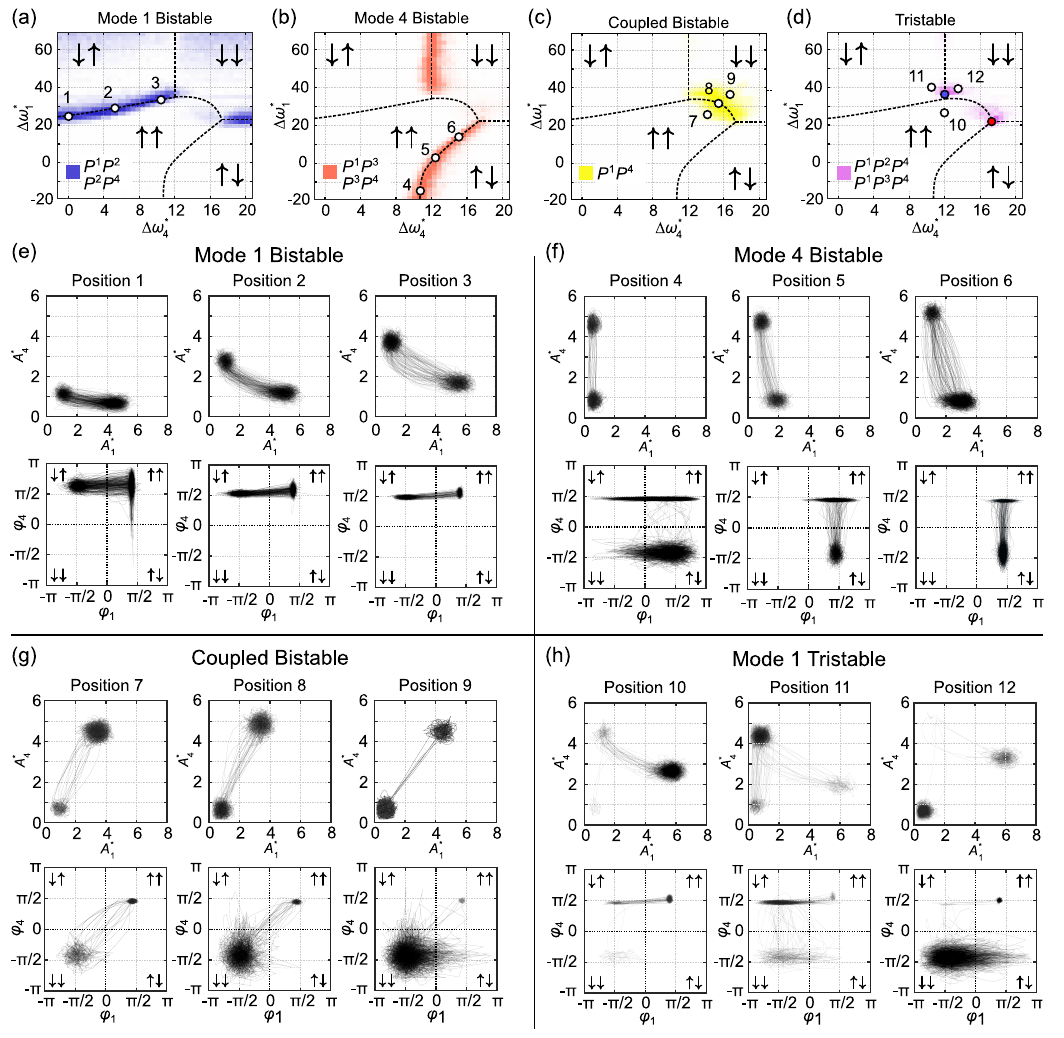}
    \caption{\label{fig:app-trend} (a-d) Joint probabilities of the coupled states as presented in the main text. (e) Lissajous plots taken from points 1-3 in Fig. \ref{fig:app-trend}a along the mode 1 single-mode KPT line. The probability of the two states remains roughly equal despite changes in relative amplitude. (f) Lissajous plots taken from points 4-6 in Fig. \ref{fig:app-trend}b along the mode 4 single-mode KPT line. The probability of the two states remains roughly equal despite changes in relative amplitude. (g) Lissajous plots taken from points 7-9 in Fig. \ref{fig:app-trend}c. Point 7 lies within the $\uparrow\uparrow$ region such that the $\uparrow\uparrow$ state is favored while point 9 lies within the $\downarrow\downarrow$ region such that the $\downarrow\downarrow$ is favored. The two state probabilities are equal at point 8, which lies on the coupled KPT line. (h) Lissajous plots taken at points 10-12 in Fig. \ref{fig:app-trend}d. Each point lies within a different stable region such that the probability of that state is dominant. 
    }
\end{figure*}

In this section, we demonstrate trends in the mode dynamics and state probabilities relative to the KPT lines to expand the discussion in the main text. Along each KPT line, the two states remain in equilibrium even as the relative amplitudes of the two modes change. We show this behavior in the two sets of Lissajous plots in Fig. \ref{fig:app-trend}e \& f. These plots are taken from three positions along the mode 1 and mode 4 single-mode bistable KPT lines that start at the left and bottom edges of Fig. \ref{fig:app-trend}a \& b respectively. Close to the edge (left-most plots), the amplitude of the monostable mode is very small. Increasing the drive frequency of the monostable mode by moving to the right or vertically, respectively, increases its amplitude and perturbs the KPT curve of the bistable mode to higher frequency. This behavior is also observed along the coupled KPT line shown in Fig. \ref{fig:app-trend}c. Away from the KPT line, as we show in Fig. \ref{fig:app-trend}g, the probability of the $\uparrow\uparrow$ and $\downarrow\downarrow$ states interchange such that the $\uparrow\uparrow$ state is favored closer to the origin and the $\downarrow\downarrow$ state is favored away from the origin. This is identical to the behavior shown for a single mode in Fig. \ref{fig:single-switching}c relative to the KPT, indicating that this is a generalized result that applies to coupled states. 

Finally, around the mode 1 triple point (blue dot in Fig. \ref{fig:kpt}h), we show three Lissajous plots taken from within the three single-state dominated regions near the mode 1 triple point, indicated as points 10-12 in Fig. \ref{fig:kpt}h. At position 10, the drive conditions are located in the $\uparrow\uparrow$ dominated region and we see that the system lingers in the $\uparrow\uparrow$ node but still undergoes transitions into the $\downarrow\uparrow$ and $\downarrow\downarrow$ states. Similarly, at positions 11 and 12 we see the system linger in the $\downarrow\uparrow$ and $\downarrow\downarrow$ states respectively. An interesting feature that can be observed at position 12 is that the $\uparrow\uparrow$ and $\downarrow\downarrow$ states are both long-lived despite the primary transition pathway between them passing through the $\downarrow\uparrow$ state. 

\bibliography{references}

\end{document}


\title{Supplementary Material for \\ ``Multistability and Noise-Induced Transitions \\in Dispersively-Coupled Nonlinear Nanomechanical Modes"}

\newcommand{\BU}{ Department of Mechanical Engineering, Division of Materials Science and Engineering, and the Photonics Center, Boston University, Boston, Massachusetts 02215, United States}

\newcommand{\VT}{Department of Mechanical Engineering, Virginia Tech, Blacksburg, Virginia 24061, United States}

\newcommand{\GC}{Department of Physics, Gordon College, Wenham, Massachusetts 01984, United States}

\newcommand{\SUNUM}{SUNUM, Nanotechnology Research and Application Center, Sabanci University, Istanbul, 34956, Turkey}

\newcommand{\Sabanci}{Faculty of Engineering and Natural Sciences, Sabanci University, Istanbul, 34956, Turkey}

\newcommand{\Bilkent}{Department of Mechanical Engineering, Bilkent University, Ankara, 06800, Turkey}

\newcommand{\UNAM}{National Nanotechnology Research Center (UNAM), Bilkent University, Ankara, 06800, Turkey}

\author{David Allemeier}
\affiliation{\BU}
 \email{davidallemeier@gmail.com}
 
 \author{I. I. Kaya}
\affiliation{\SUNUM}
\affiliation{\Sabanci}

\author{M. S. Hanay}
\affiliation{\Bilkent}
\affiliation{\UNAM}

\author{K. L. Ekinci}
 \email{ekinci@bu.edu}
\affiliation{\BU}

\date{\today}

\maketitle
\widetext
\tableofcontents

\newpage

\setcounter{equation}{0}
\setcounter{figure}{0}
\setcounter{table}{0}
\setcounter{page}{1}
\makeatletter
\renewcommand{\theequation}{S\arabic{equation}}
\renewcommand{\thefigure}{S\arabic{figure}}
\renewcommand{\thetable}{S\arabic{table}}
\renewcommand{\bibnumfmt}[1]{[S#1]}
\renewcommand{\citenumfont}[1]{S#1}

\maketitle

\label{section:SI_Experimental_Details}
\section{Calibration of the Mode Amplitudes}
\label{sec:amp}
\begin{figure*}[b]
    \includegraphics[]{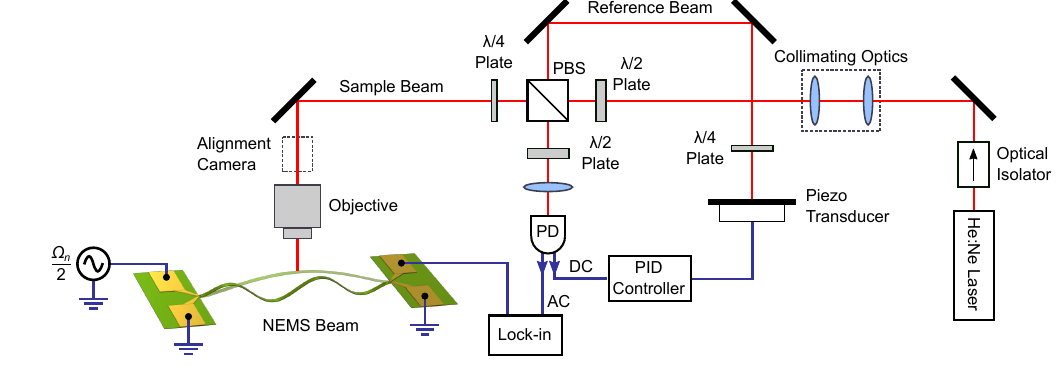}
    \caption{\label{fig:si-interferometer} Schematic of the path-stabilized homodyne Michelson interferometer used to calibrate the electrical measurements. We prepare a 633 nm He:Ne laser source using an optical isolator and collimator and rotate its polarization with a half-wave plate. The source beam is split into reference and sample beams using a polarizing beam splitter (PBS). The sample beam, on the P-channel, is passed through a quarter-wave plate and focused onto the NEMS beam using a 100$\times$/0.5 NA objective lens. The reference beam, on the S-channel, is directed to a piezo-actuated mirror through a quarter-wave plate. The two reflected beams are recombined in the PBS and passed through a half-wave plate to generate an interference signal before detection in the photodiode (PD). The output signal from the PD consists of an AC ($>$ 25 kHz) and DC ($<$ 25 kHz) component. The DC component is used for path stabilization by compensating for slow drift in the reference arm path length via a PID controller and the piezo-actuated mirror. This keeps the interferometer locked to the center of the interference signal and ensures the signal remains linear over the maximum range of displacements \cite{steen_nl_inter}. The AC component is fed to the lock-in amplifier to detect the beam motion. The NEMS device was actuated with a sinusoidal drive at half its eigenfrequency using one electrode; the other electrode was used for piezoresistive detection using a DC bias of 100 mV.
    }
\end{figure*}

This experiment utilized two different setups. The primary results were obtained from piezoresistance measurements using the circuit shown in Fig. 2 and described in the main text. Piezoresistance provides a highly linear, stable, and weakly mode-dependent responsivity at large displacement \cite{matheny_thesis, chaoyang}. To convert the piezoresistance measurements from volts to nanometers,  we employed a path-stabilized homodyne Michelson interferometer, illustrated in Fig. \ref{fig:si-interferometer}, that utilized a He:Ne laser ($\lambda \approx 633$ nm). The interferometer  provides a highly sensitive absolute measurement of the displacement. The measured amplitude depends strongly on the proximity of the optical spot to a mode anti-node and is linear only at relatively small displacements. Here,  we find that the linear approximation generates an error of 5\% at a displacement of 20 nm ($\lambda/32$). This nonlinearity is predictable and can be corrected with a calibration curve \cite{wagner_interferometer, steen_nl_inter}.

\begin{figure}
    \centering
    \includegraphics[]{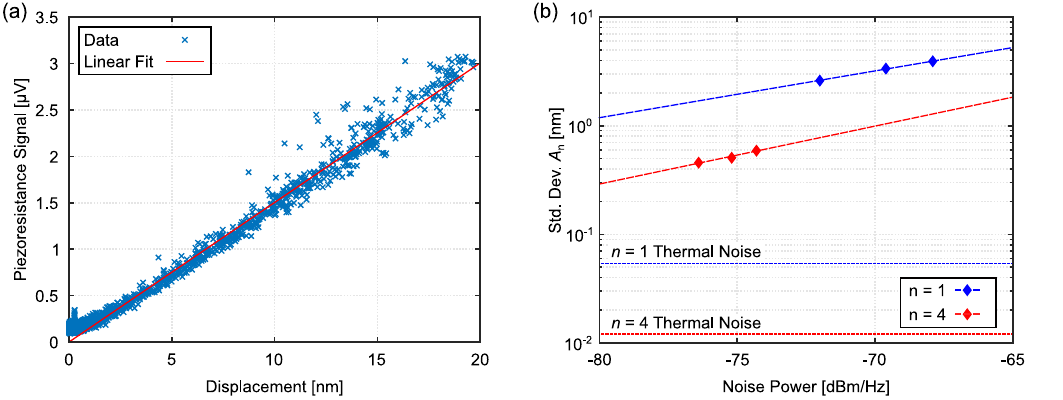}
    \caption{(a) Calibration of the piezoresistance measurements using a Michelson interferometer. Data combines seven forward frequency sweeps of mode 1 at drive powers between -27 dBm and -15 dBm. Thermal noise amplitude is estimated to be 38 pm for mode 1. (b) Measured rms noise amplitude of modes $n = 1$ and 4 measured at $\Delta \omega_n^* = 0$ at three different noise power spectral densities. Dashed lines are linear fits. The estimated thermal noise floor for both modes is shown with faint lines. The estimated thermal noise floor values are 54 pm and 12 pm for modes $n=1$ and 4, respectively, calculated using $\left<A_n\right> = \sqrt{k_BT/k_n}$ at $T = 300$ K, $k_B = 1.38 \times 10^{-23}$ J/K and the modal spring constants $k_1 = 1.4$ N/m and $k_4 = 26.6$ N/m.
    \label{fig:si-responsivity}}
\end{figure}

To calibrate the displacement, the  laser was focused at the midpoint of the NEMS beam; this corresponds to the anti-node of the first mode. The NEMS beam was actuated on one side with sinusoidal drives between -27 dBm and -15 dBm to produce a maximum displacement of 20 nm, within the linear regime of the interferometer and sufficient to reach the nonlinear mechanical regime. For each drive power we conducted a frequency sweep of the first mode, while simultaneously recording the piezoresistance and optical signals using a lock-in amplifier. We plot these data  in Figure \ref{fig:si-responsivity}a and find that the piezoresistance signal is highly linear and shows a saturation at low amplitude ($\lesssim1$ nm), possibly  due to the higher background noise  of the piezoelectric transducer. We repeated the experiment for the fourth mode by focusing the laser at 3/8 of the beam length, corresponding to one of the anti-nodes. Linear regression provided output transducer responsivities $\xi_n$ of 6.12 nm/$\mu$V and 1.80 nm/$\mu$V for the first and fourth modes, respectively.  We used these $\xi_n$ values to determine the displacement  from the measured piezoresistance signal in all experiments. The transducer responsivities are listed in     Table \ref{tab:beam_params}.

Based on this calibration, we found that the maximum displacement in our experiment was approximately $250$ nm. To ensure the accuracy of the calibration at this displacement, we now estimate the nonlinearity in the piezoresistance measurements. The nonlinearity in the piezoresistive detection results from changes in the electrical resistance increase due to elongation, which is a geometric effect, and oscillatory heating \cite{chaoyang, monan_scaling}. As discussed in the Supplementary Information of \cite{monan_scaling}, we can expand the base resistance $R_0$ to include higher order terms. The first order geometric correction to the resistance, $R^{(1)}_p$, is related to $R_0$ according to \cite{monan_scaling},
\begin{equation}
    \frac{R^{(1)}_p}{R_0} = \frac{\gamma {A_n}^2\int_0^{l_r} \left[\partial\Phi_n(z) \over \partial z\right]^2 dz}{2l_r},
\end{equation}
which depends on the gauge factor $\gamma \approx 10$ \cite{chaoyang}, the length of the resistor $l_r \approx 600$ nm, the modal amplitude $A_n$, and the mode shape $\Phi_n(z)$. By approximating the eigenmodes as those of a string, $\Phi_n(z) = \sin{({n\pi z \over L})}$, we find the error at the maximum experimental displacements of $A_1 \approx 250$ nm and $A_4 \approx 35$ nm to be 5\% and 1.5\%, respectively. The geometric error varies quadratically with the mode amplitude; at the maximum mode amplitudes in the mapping experiments, $A_1 = 110$ nm and $A_4 = 20$ nm, the error is 1\% and 0.5\%, respectively. Similarly, the first order correction to resistance due to heating was found to be $<5\%$ in functionally identical beams \cite{monan_scaling} at the highest drive voltages.

\section{Input Noise}
\label{sec:noise}

The transition behavior of a bistable NEMS device has been shown to depend on the characteristics of the noise \cite{chan_poisson}. We carefully crafted and characterized our applied noise to avoid unwanted effects. As depicted in Fig. 2 in the main text, we started with two broadband Gaussian noise signals generated by an arbitrary wave generator (AWG). Both broadband Gaussian noise signals were low-pass filtered at 50 kHz and high-pass filtered at 10 kHz, then amplified by a factor of 100. The two signals were then mixed with two separate 1 V sinusoidal local oscillator (LO) signals offset from half the selected mode eigenfrequencies by 20 kHz. The up-converted noise of the first mode was additionally low-pass filtered at 1.9 MHz. Finally, the two up-converted bandpass noise signals were superposed with their respective sinusoidal drives within the signal adder inside the lock-in amplifier. 

\begin{figure*}
    \includegraphics[]{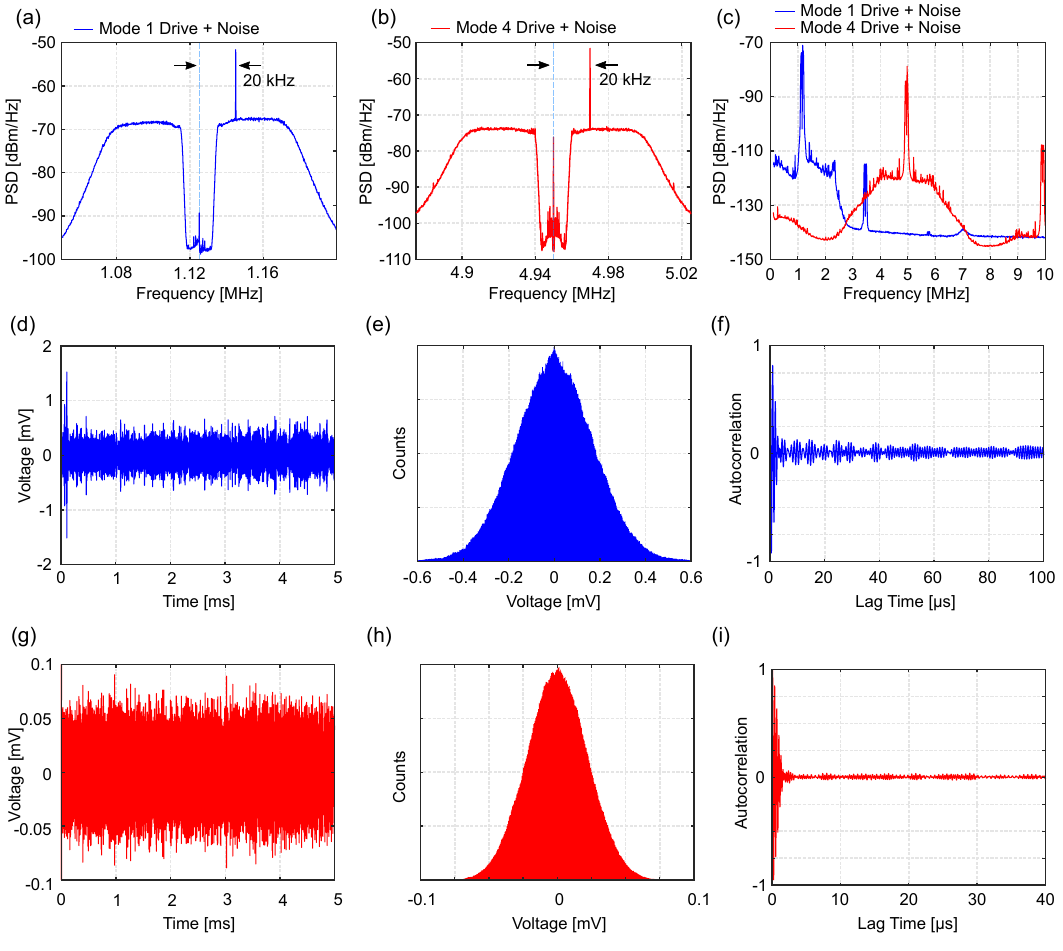}
    \caption{\label{fig:si-signal} (a \& b) Power spectral density of the drive signal for the first and fourth mode, including the sinusoidal drive and narrowband noise. The average noise power spectral density within the mode linewidth is -67.9 dBm/Hz and -74.2 dBm/Hz for modes 1 and 4, respectively. (c) Wide-band power spectra of the first and fourth mode drives. Spectra were recorded by measuring the drive signal in a spectrum analyzer. (d \& g) Time-domain measurement of the noise signal for the first and fourth modes. (e \& h) Histogram of the noise signal of modes 1 and 4, respectively, showing a  Gaussian distribution. (f \& i) Autocorrelation of the noise signals showing a coherence time of $<5$ $\mu$s.
    }
\end{figure*}

In Fig. \ref{fig:si-signal}a, b and c, we show the power spectral density (PSD) of the two drive signals measured using a spectrum analyzer. Near the mode eigenfrequencies in Figs. \ref{fig:si-signal}a \& b, we observe the sinusoidal drive superposed on two narrowband noise signals, one of which overlaps with the drive signal. Between the two mixed bands we observe a small leakage from  the mixers. More specifically, for the first mode (Fig. \ref{fig:si-signal}a), the  eigenfrequency is ${\Omega_1 \over 2 \pi} \approx 2.292$ MHz, with ${\Omega_1 \over 4 \pi} \approx 1.146$ MHz. The sinusoidal drive peak can be seen at 1.146 MHz. The narrowband noise can be seen to extend roughly 10 kHz below and 30 kHz above the drive peak. The mode noise bandwidth is  ${\pi (\Omega_1/2\pi) \over 2  Q_1} \approx 180 $ Hz. Similar observations can be made in Fig. Fig. \ref{fig:si-signal}b, where $\Omega_4 \over 2\pi$ is 9.995 MHz and the mode noise bandwidth is approximately 1.5 kHz.  When observed in a wider frequency span in Fig. \ref{fig:si-signal}c, we see that the harmonics of the mixed noise signal have very low amplitude. 

The applied electrical noise signal drives the beam stochastically and produces mechanical noise that we detect in the motion of the beam. Returning to Fig. S2b, we show the measured noise amplitudes of the beam in modes 1 and 4 in nm as a function of the input noise power spectral densities with fixed coherent  sinusoidal drive powers of $ -10$ dBm ($F_1^* = 40$) and $ -6.8$ dBm ($F_4^* = 11.7$), i.e., the drive conditions for  the stability mapping experiment shown in Fig. 11.  (The data shown in Fig. 11 corresponds to the highest noise power spectral density of $D_1=-72~ {\rm dBm/Hz},~ D_4 = -76.4~\rm dBm/Hz$.) In Fig. S2b, the coherent drive signal was applied exactly at the eigenfrequency and the detection was performed using a lock-in detection bandwidth of $\sim 300$ Hz centered at the eigenfrequency. The measured fluctuation amplitude was found to increase monotonically with the applied noise power. In \S S:\ref{sec:force} below, we will compare the expected and measured noise amplitudes.

Finally, we analyzed the applied noise by recording the signals in the time domain over a period of 5 ms with a sample rate of 2 GHz. We extracted the frequency components near the beam resonance using a band-pass filter function in MATLAB and measured the autocorrelation and histogram. The results for the first and fourth modes are shown in Figs. \ref{fig:si-signal}d-e and f-h, respectively. We found that the histograms in Figs. \ref{fig:si-signal}e \& h show a clear Gaussian distribution; the autocorrelation in Figs. \ref{fig:si-signal}f \& i show correlation times of 3 $\mu s$ and 1.5 $\mu s$ respectively.

\section{Measurement of the Duffing Coefficients and Bifurcation Frequencies}
\label{sec:duffing}

\begin{figure*}[]
    \includegraphics[]{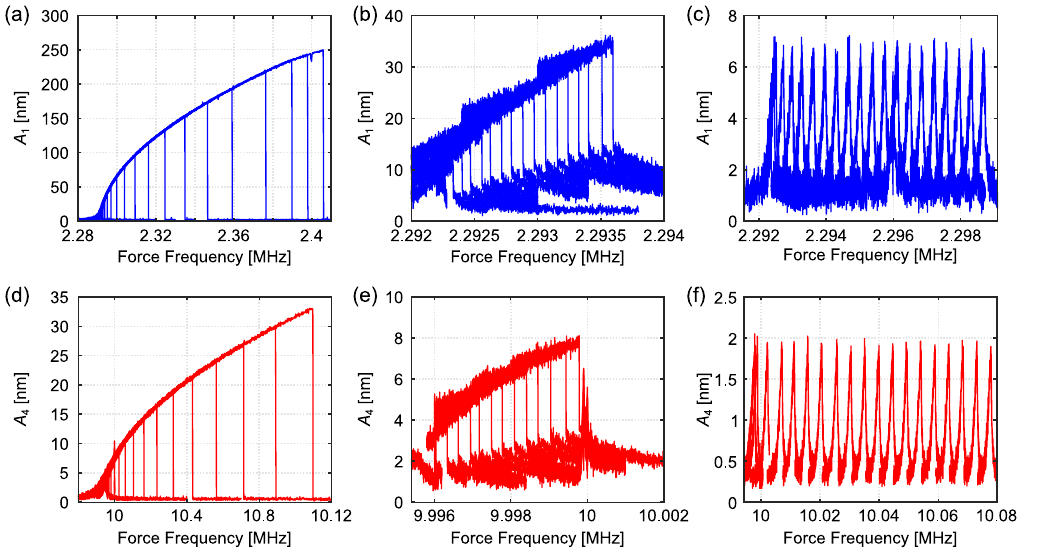}
    \caption{\label{fig:si-duffing} (a \& d) Measurement of the Duffing backbone and the upper bifurcation points using a series of forward frequency sweeps at varied drive power. Mode 1 was driven with drive powers between -23.5 dBm and -7 dBm; mode 4 was driven with powers between -16.8 dBm and -4.8 dBm. The force frequency along the $x$-axis is twice the electrical drive frequency. (b \& e) Measurement of the lower bifurcation points at the same drive powers using reverse frequency sweeps. (c \& f) Forward frequency sweeps of a weakly-driven mode at varied amplitudes of a strongly-mode. The amplitude of the strongly-driven mode was controlled by varying the drive frequency along its Duffing backbone. For part c, mode 1 was weakly-driven at -23.5 dBm while mode 4 was strongly-driven at -4.8 dBm; in part d, mode 1 was strongly-driven at -7 dBm while mode 4 was weakly-driven at -16.8 dBm. The maximum amplitudes of the strongly-driven modes were $A_1 = 65$ nm and $A_4 = 38$ nm.
    }
\end{figure*}

The intramodal Duffing coefficients were calculated from a series of experimental forward frequency sweeps of each mode at varied drive powers between -27 and -4.8 dBm. In these experiments, shown in Fig. \ref{fig:si-duffing}a \& d, we used piezoresistive detection and actuation as described in the main text, where we drove each mode of the beam with an AC signal on one side and detected on the other through a DC bias of 100 mV. To compensate for the changing drive power of the mode-under-test, which leads to thermal drift in the eigenfrequency, we added an additional AC signal off-resonance and away from combination frequencies to maintain constant power \cite{monan_scaling}. The drive and detection was conducted with a lock-in amplifier. The amplitude is expressed in nanometers using the conversion coefficients obtained from the Michelson interferometer discussed in \S S:\ref{sec:amp}. We recorded the frequency and amplitude at the upper bifurcation point, where the amplitude drops to the lower branch, and conducted a linear fit of $\omega_{nU}$ versus ${A_{nU}}^2$ in Fig. \ref{fig:si-fitting}a \& e. The slope and intercept provided $\alpha_{nn}$ and $\Omega_n$, respectively. From the linear fit, we obtained intramodal Duffing coefficients of $\alpha_{11}/4\pi^2 = 1.08 \times 10^{-5}$ MHz$^2$/nm$^2$ and $\alpha_{44}/4\pi^2 = 277.1 \times 10^{-5}$ MHz$^2$/nm$^2$. The theoretical Duffing coefficient of a string under tension can be expressed in terms of the physical constants of the system as \cite{monan_scaling,lc_nems},
\begin{equation}
    \label{eq:si-alpha}
    \alpha_{nm} = \frac{E\pi^4}{4\rho L^4} n^2 m^2.
\end{equation}
Using this equation, we calculate theoretical Duffing coefficients of $\alpha_{11}/4\pi^2 = 0.8 \times 10^{-5}$ MHz$^2$/nm$^2$ and $\alpha_{44}/4\pi^2 = 210 \times 10^{-5}$ MHz$^2$/nm$^2$.

For each drive power, we measured the lower bifurcation point using a separate reverse frequency sweep. These measurements were conducted immediately after the forward frequency sweeps shown in Fig. \ref{fig:si-duffing}a \& d and used higher frequency resolution. Before each measurement we ensured that the mode was on its lower branch by changing the drive frequency to far above the eigenfrequency. We then swept the drive frequency downward until the mode jumped to its upper branch. We recorded the frequency of the lower bifurcation point for each force and used the values to calibrate the force and quality factor as discussed in \S S:\ref{sec:force} below. The reverse frequency sweeps are presented in Fig. \ref{fig:si-duffing}b \& e in units of nanometers using the conversion coefficients found above. Both the upper and lower bifurcation frequencies are plotted in Fig. 3b in the main text as a function of the drive force.

Similarly, as described in the main text and reproduced here in Fig. \ref{fig:si-duffing}c \& f, the intermodal Duffing coefficients were calculated using a series of forward frequency sweeps of a weakly-driven mode (mode $n$) in the linear regime at varied amplitudes of a strongly-driven nonlinear mode (mode $k$). The two modes were driven and detected simultaneously on opposite sides of the beam, as described in the main text. The amplitude of the strongly-driven mode ($A_k$) was controlled by varying the drive detuning frequency $\Delta \omega_k \over 2\pi$ with a constant drive power to move it along its Duffing backbone. Through Eq. 7, this produces a constant step in ${A_k}^2$ for a constant step in $\Delta \omega_k$. Accordingly, this approach also produces a constant intermodal angular eigenfrequency shift $\Delta \Omega_{nk}$ that results in the evenly spaced peaks. A linear fit of the drive frequency at each peak of mode $n$ versus ${A_k}^2$ provides the intermodal Duffing coefficients using Eq. 14, shown in Fig. \ref{fig:si-fitting}b \& f. The experimental values obtained were $\alpha_{14}/4\pi^2 = 17.7 \times 10^{-5}$ MHz$^2$/nm$^2$ and $\alpha_{41}/4\pi^2 = 17.8 \times 10^{-5}$ MHz$^2$/nm$^2$. Using Eq. \ref{eq:si-alpha}, we calculate theoretical values of $\alpha_{14}/4\pi^2 = \alpha_{41}/4\pi^2 = 13 \times 10^{-5}$ MHz$^2$/nm$^2$.

\section{Calibration of the Force (Input) Transducer and Modal Quality Factors}
\label{sec:force}

The drive force is only known  in terms of the applied voltage. In this section, we calibrate the drive force $F_n$ and quality factor $Q_n$ using the measured angular frequencies of the upper and lower bifurcation points, $\Delta \omega_{nU}$ and $\Delta \omega_{nL}$,  respectively (Fig. \ref{fig:si-duffing}a, b, d, and e). First, we assume that the modal force $F_n$, in units of Newtons, is related to the applied voltage according to $F_n = \kappa_n {V_n}^2$, where $\kappa_n$ is the input transducer responsivity specific to mode $n$ with units of N/mV$^2$. The dimensionless force $F_n^*$ can then be written as,
\begin{equation}
    F^*_n = {\sqrt{\alpha_{nn}{Q_n}^3} \over {\Omega_n}^3 M_n}\kappa_n{V_n}^2,
\end{equation}
which is related to the intramodal Duffing coefficient $\alpha_{nn}$, angular eigenfrequency $\Omega_n$, the modal mass $M_n$, and the quality factor $Q_n$. Care must be taken to align the units properly---i.e., kilograms, meters, and rad/s. The eigenfrequency and Duffing coefficient were found experimentally in \S S:\ref{sec:duffing}; we now seek the quality factor $Q_n$ and  the input transducer responsivity $\kappa_n$  to close the equation.

\begin{figure*}
    \includegraphics[]{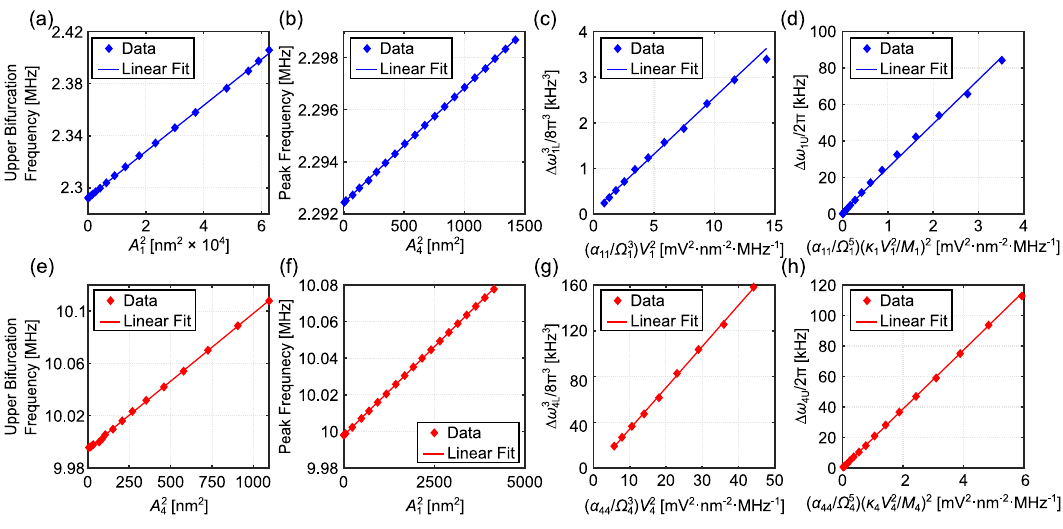}
    \caption{\label{fig:si-fitting} (a \& e) Calculation of $\alpha_{nn}$ and $\Omega_n$ from the upper bifurcation points recorded from the forward frequency sweeps shown in Fig. \ref{fig:si-duffing}a \& d. (b \& f) Calculation of $\alpha_{nk}$ from the peak frequency $\Delta \omega_{n,\rm max}$ of a weakly-driven mode $n$ at varied amplitudes of a strongly-driven mode $k$. Sweep data are shown in Fig. \ref{fig:si-duffing}c \& f. (c \& g) Fitting of the lower bifurcation frequencies to determine the force calibration coefficients ${\kappa_n}$. (d \& h) Fitting of the upper bifurcation frequencies to calculate the quality factor.
    }
\end{figure*}

Using the relations in the main text, we can non-dimensionalize the (angular) force detuning frequency $\Delta \omega_n$, which is expressed relative to the unperturbed (angular) eigenfrequency $\Omega_n$, as
\[ \Delta \omega_n^* = {Q_n \over \Omega_n}\Delta \omega_n.\]
We note that, with this nondimensionalization, angular frequency and frequency become the same parameter. This allows us to write the dimensionless bifurcation frequencies as $\Delta \omega_{nL}^* = {Q_n \over \Omega_n}\Delta \omega_{nL}$ and $\Delta \omega_{nU}^* = {Q_n \over \Omega_n}\Delta \omega_{nU}$. The dimensionless force $F_n^*$ is directly related to the dimensionless bifurcation frequencies $\Delta \omega_{nU}^*$ and $\Delta \omega_{nL}^*$ at large detuning through the asymptotic relations
\begin{eqnarray}
        \label{eq:bif_approx}
        \Delta \omega^*_L &=& \sqrt[3]{\frac{81}{128}}{F^*}^{2/3}, \\ \Delta \omega^*_U &=& \frac{3}{8}{F^*}^2,
\end{eqnarray}
which are derived in the Appendix. To calibrate the force, we first consider the lower bifurcation point. Inserting the  expressions for $F_n^*$ and $\Delta \omega_{nL}^*$ into the equation for the lower bifurcation, we find a relation between the applied voltage and the angular force detuning at the lower bifurcation,
\begin{equation}
    {\Delta \omega_{nL}}^3 = {81 \over 128}{\alpha_{nn} \over {\Omega_n}^3}{{\kappa_{n}}^2 \over {M_n}^2} {V_n}^4,
\end{equation}
which no longer depends on $Q_n$ and depends only on the unknown constant ${\kappa_{n}}^2/{M_n}^2$.  Using the measured lower bifurcation frequencies from  Fig. \ref{fig:si-duffing}b \& e, we plotted the quantity ${\alpha_{nn} \over {\Omega_n}^3} {V_n}^4$ versus the bifurcation frequency detuning cubed, $\left(\Delta \omega_{nL} \over 2 \pi\right)^3$. We then  calculated a linear fit to obtain ${{\kappa_{n}}^2 \over M_n^2}$; these fits are shown in Figs. \ref{fig:si-fitting}c \& g. We  approximated the modal mass as $M_n = 6.8$ pg from $M_n=\rho LS/2$ and used the calculated slope to find $\kappa_1=2.1 \times 10^{-15}$ N/mV$^2$ and $\kappa_4=8.0 \times 10^{-15}$ N/mV$^2$. These values are listed in Table \ref{tab:beam_params}.

Next, we extracted the upper bifurcation points from Fig. \ref{fig:si-duffing}a \& d to now calculate $Q_n$. From the above relations, the expression for the angular frequency detuning at the upper bifurcation can be written as,
\begin{equation}
    \Delta \omega_{nU} = {3 \over 8}{\alpha_{nn} \over {\Omega_n}^5}\left({{\kappa_n}^2 \over {M_n}^2}\right){Q_n}^2{V_n}^4,
\end{equation}
which includes the term ${{\kappa_n}^2 \over M_n^2}$ that we found using the lower bifurcation data. All terms are now known except $Q_n$, so we plot the upper bifurcation detuning $\Delta \omega_{nU}$ against the quantity ${\alpha_{nn} \over {\Omega_n}^5}\left({{\kappa_n}^2 \over {M_n}^2}\right){V_n}^4$. The square root of the slope yields the quality factors of $Q_1 = 20,100$ and $Q_4 = 11,000$; these fits are shown in Fig. \ref{fig:si-fitting}d \& h.

\section{Estimation of Beam Amplitude from Transducer Responsivities}

\subsection{Estimation of Coherent Beam Amplitude from Drive Voltage}

To validate the calibration, we can compare the measured and expected displacements for a given drive voltage in the linear regime using the linear susceptibility relation,
\begin{equation}
    A_n = \frac{F_nQ_n}{k_n} = \frac{\kappa_n{V_n}^2 Q_n}{k_n}.
\end{equation}
At $V_1 = 15$ mV, the displacement of mode 1 was measured to be $A_1\approx 7.5$ nm, which can be seen in the smallest-amplitude frequency sweep in Fig. \ref{fig:si-duffing}b. Using the values of $\kappa_1 = 2.1 \times 10^{-15}$ N/mV$^2$, $Q_1 = 20,100$, and $k_1 = 1.4$ N/m, we obtain an expected displacement of 6.8 nm, which is close to the calculated value. For mode 4 driven at 25 mV, we observed a maximum displacement of about 2 nm. Using values of $\kappa_4 = 8.0 \times 10^{-15}$ N/mV$^2$, $Q_4 = 11,000$, and $k_4 = 26.6$ N/m, the predicted displacement is 2.1 nm. These conversion factors were used to calculate the amplitude and forcing at the critical dimensionless theory values of $A^*_c = \left(4/3\right)^{3/4}$ and $F^*_c = \left(4/3\right)^{5/4}$, giving values of $A_{1c} = 6.1$ nm, $A_{4c} = 2.2$ nm, $F_{1c} = 0.5$ pN, and $F_{4c} = 6.3$ pN \cite{lc_nems, nayfeh_nonlinear}. The results match the experimentally observed values shown in Figs. 3b and \ref{fig:si-duffing}b \& e.

\begin{table*}[t]
    \begin{tabular}{c|c|c|}
        Mode $n$ & ${\xi_n}$ [nm/$\mu$V] & $\kappa_n$ [$\rm N/mV^2$]   \\ [0.5ex] \hline \hline 1 & 6.12 & $2.1 \times 10^{-15}$  \\ 
        4 & 1.8 & $8.0 \times 10^{-15} $
    \end{tabular}

    \caption{Values of output ($\xi_n$) and input ($\kappa_n$) transducer responsivities. }
    \label{tab:beam_params}
\end{table*}

\subsection{Estimation of Noise Amplitude from Input Noise Voltage}

In order to apply this calibration to estimate the noise amplitude of the beam, we must first determine the effective  force noise on the beam. As discussed in the previous section, the mechanical force is related to the applied voltage according to $F_n = \kappa_n{V_n}^2$ with the voltage $V_n$ being a sinusoidal. The applied voltage here, however, is a superposition of 1) a sinusoidal drive at the angular frequency $\Omega_n/2$ and amplitude $V_n$, and 2) a narrowband noise signal centered around $\Omega_n/2$. Following Robins \cite{robins_txt}, we can express the total input voltage as,
\begin{equation} \label{eq:noise_sum}
    V_{in}(t)=V_n\cos\left({\Omega_n t \over 2}\right) + \sum\limits_k \sqrt{2D_n} {\cos \left[ {\left( {{{{\Omega _n}} \over 2} + 2\pi k} \right)t + {\varphi _k}} \right]},
\end{equation}
where $D_n$ is average noise power in 1 Hz, i.e., noise power spectral density, and is assumed to remain flat over frequency (Fig. \ref{fig:si-signal}). The individual spectral lines in the noise voltage are of bandwidth 1 Hz and are offset from the carrier by $\pm k$ Hz; the sum accounts for the total bandwidth and noise of the narrowband signal. The phases of individual sinusoidals $\varphi_k$ are assumed to be random and slowly varying variables. 

The total force is proportional to the square of $V_{in}(t)$. If $V_{in}(t)$ is squared, one obtains, to leading order, a coherent signal at angular frequency $\Omega_n$ as before and a fluctuating signal $\sum\limits_k {{{{V_n}{\sqrt D_n}} \over {\sqrt 2 }}\cos \left[ {\left( {{\Omega _n} + 2\pi k} \right)t + {\varphi _k}} \right]} $, which is a narrowband noise signal now centered at $\Omega_n$. The last term in the expansion of the square is ignored, given that $V_n \gg \sqrt{2D_n}$. Thus, if the input voltage noise has a Gaussian distribution, the force noise will also have an  approximately Gaussian distribution.  The square of the first term in  Eq. \ref{eq:noise_sum} will be deterministic, and the small  non-Gaussian term, i.e., the square of the noise term in  Eq. \ref{eq:noise_sum}, is ignored. The mechanical mode itself is a sharp (high $Q$) filter that only responds to fluctuating drives that are within the noise bandwidth of the mode ${\pi \over 2}\times{2\Gamma_n \over 2\pi}$. We can thus write the total force noise acting on the mode as
\begin{equation}
    F_{N_n}\approx \kappa_n V_n \sqrt{D_n} \sqrt{\pi \over 2} \sqrt{2\Gamma_n \over 2\pi}.
\end{equation}

In the first mode, with a noise power spectral density of $D_1 = -68$ dBm/Hz and a noise bandwidth of ${\pi \over 2}\times{2\Gamma_1 \over 2\pi}=180$ Hz, we find that the equivalent peak voltage noise amplitude is $\sim 1.8$ mV. Using the calibrated force conversion factor of $\kappa_1 = 2.1 \times 10^{-15}$ N/mV$^2$ and a peak drive voltage of $V_1 = 70$ mV, we convert this into a  force noise of $F_{N_1} = 0.26$ pN. Since this is a sinusoidal for all practical purposes due to the high $Q$, we use the linear susceptibility approximation to find the  noise amplitude 
\begin{equation}
    A_{N_n}\approx \frac{F_{N_n} Q_n}{k_n}.
\end{equation}
We find an estimate of approximately 4 nm peak and 2.9 nm rms noise amplitude---which is not far from the measured peak and rms values of 5.6 nm and 4 nm, respectively, reported in Fig. \ref{fig:si-responsivity}b. For the fourth mode, we predict a peak noise amplitude of 0.5 nm (0.4 nm rms), compared to a measured peak noise amplitude of 0.8 nm (0.6 nm rms); we note that the measured noise in the fourth mode is an underestimate based on the relative bandwidth of the lock-in ($\sim 300$ Hz) and the noise bandwidth of mode 4 ($\sim 1.5$ kHz).

We note that the electrical noise from detection should increase the measured noise. Also, in the measurements in Figs. 7-11, the lock-in bandwidth and the natural noise bandwidth of the mode are not perfectly aligned, resulting in only a portion of the available mechanical noise being admitted into the lock-in. Finally, the nonlinear higher-order noise components resulting from the electrothermal actuation are entirely neglected in our analysis.

\bibliography{references}